\documentclass[12pt]{article}

\usepackage{amsmath,amsthm, amssymb,amsfonts,bbm}
\usepackage{epsfig,cancel}

\numberwithin{equation}{section}

\newcommand{\mycaption}[1]{\caption{{\sf \small #1}}}

\newcommand{\beq}{\begin{equation}}
\newcommand{\eeq}{\end{equation}}
\newcommand{\ba}{\begin{array}}
\newcommand{\ea}{\end{array}}
\newcommand{\bea}{\begin{eqnarray}}
\newcommand{\eea}{\end{eqnarray}}
\newcommand{\bean}{\begin{eqnarray*}}
\newcommand{\eean}{\end{eqnarray*}}
\newcommand{\eref}[1]{(\ref{#1})}

\newcommand{\comment}[1]{}

\newcommand{\cO}{{\cal O}}
\newcommand{\cN}{{\cal N}}
\newcommand{\cF}{{\cal F}}
\newcommand{\cA}{{\cal A}}
\newcommand{\cB}{{\cal B}}
\newcommand{\cC}{{\cal C}}

\newcommand{\cV}{{\cal V}}

\def\cjn1{{\cA, \cC^*\otimes \wedge^j \cN^*}}
\def\bjn1{{\cA, \cB^*\otimes \wedge^j \cN^*}}
\def\vjn1{{\cA, \cV^*\otimes \wedge^j \cN^*}}
\def\cjn2{{\cA, \cC\otimes \wedge^j \cN^*}}
\def\bjn2{{\cA, \cB\otimes \wedge^j \cN^*}}
\def\vjn2{{\cA, \cV\otimes \wedge^j \cN^*}}

\def\fnote#1#2{\begingroup\def\thefootnote{#1}\footnote{#2}
     \addtocounter{footnote}{-1}\endgroup}

\begin{document}

\vspace{1cm}

\title{{\huge \bf
Yukawa Textures From Heterotic Stability Walls
}}

\vspace{2cm}

\author{
Lara B. Anderson${}^1$, James Gray ${}^2$, and Burt Ovrut${}^{1}$
}
\date{}
\maketitle
\begin{center} {\small ${}^1${\it Department of Physics, University of
      Pennsylvania, \\ Philadelphia, PA 19104-6395, U.S.A.}
    \\${}^2${\it Rudolf Peierls Centre for Theoretical Physics, Oxford
      University,\\
      $~~~~~$ 1 Keble Road, Oxford, OX1 3NP, U.K.}\\
    \fnote{}{andlara@physics.upenn.edu, j.gray1@physics.ox.ac.uk, ovrut@elcapitan.hep.upenn.edu} }
\end{center}

\abstract{\noindent A holomorphic vector bundle on a Calabi-Yau
  threefold, $X$, with $h^{1,1}(X)\geq 2$ can have regions of its K\"ahler cone
  where it is slope-stable, that is, where the four-dimensional theory
  is ${\cal{N}}=1$ supersymmetric, bounded by ``walls of
  stability''. On these walls the bundle becomes poly-stable, decomposing into a direct sum, and the low  
  energy gauge group is enhanced by at least one anomalous
  $U(1)$ gauge factor. In this paper, we show that these additional
  symmetries can strongly constrain the superpotential in the stable
  region, leading to non-trivial textures of Yukawa interactions and
  restrictions on allowed masses for vector-like pairs of matter
  multiplets. The Yukawa textures exhibit a hierarchy; large couplings arise on the stability wall 
  and some suppressed interactions ``grow back'' off the wall, where the extended $U(1)$ symmetries are spontaneously broken. A number of explicit examples are presented involving
  both one and two stability walls, with different decompositions of
  the bundle structure group. A three family standard-like model with
  no vector-like pairs is given as an example of a class of $SU(4)$
  bundles that has a naturally heavy third quark/lepton family.
  Finally, we present the complete set of Yukawa textures that can arise for any
  holomorphic bundle with one stability wall where the structure
  group breaks into two factors.}

\newpage

\tableofcontents

%
%

\section{Introduction}
Compactifications of heterotic
string and M-theory \cite{Candelas:1985en}-\cite{Gray:2007zza} on smooth Calabi-Yau threefolds are an important 
approach to string phenomenology \cite{burt}. In several recent
papers \cite{Anderson:2009sw,Anderson:2009nt}, the phenomenon of
{\it stability walls} was explored within this context. The idea 
behind these structures is simple. Heterotic compactifications necessarily
involve background gauge fields on the
Calabi-Yau space. These are normally chosen so as to
preserve ${\cal N}=1$ supersymmetry in four dimensions. Hence, they must satisfy the the Hermitian Yang-Mills equations with zero slope, equations notoriously difficult to solve. What was shown in 
\cite{Anderson:2009sw,Anderson:2009nt} is that even if a solution is found in some regions of K\"ahler moduli space, there is not, in general, a solution in other regions.
On the boundary between the ``chambers'' of K\"ahler moduli
space where supersymmetry is or is not preserved, co-dimension one ``stability
walls'' appear. 

On these walls, new Abelian gauge bosons
become light and the gauge symmetry of the associated four-dimensional effective
theory is enhanced. Although these additional $U(1)$ symmetries are spontaneously broken in the interior of a supersymmetric region, their effect continues to be felt. In particular, 
matter fields and some moduli
have specific charges under the enhanced
symmetries. 
These charges
restrict the form of operators which can appear in the four-dimensional superpotential, 
not simply on or near the stability wall
but, via holomorphy arguments, throughout the entire supersymmetric region.
In this paper, we describe the textures in Yukawa
couplings 
that can result from the
presence of stability walls in the K\"ahler cone. 
We also analyze the constraints these walls can impose on the masses of vector-like pairs of matter multiplets.
This is useful for both
bottom-up and top-down approaches to phenomenology. From the
bottom-up point of view, our analysis will provide a broad and
well-defined set of Yukawa and vector-like pair mass textures that can arise naturally in
smooth compactifications of heterotic string and M-theory. These
textures can be used in model building, and can act
as a guide as to what is likely to occur in the heterotic context. In particular, in Appendix A we list all textures that can result from the simplest kinds of
stability walls.

From the top-down perspective, our results can also act as an guide to model building. Finding stability
wall structure is relatively straightforward within the context of holomorphic vector bundle  
constructions such as monads \cite{Distler:1987ee}-\cite{Blumenhagen:2006ux}, extensions \cite{burt,AG,Braun:2005zv,Braun:2004xv} and
spectral covers \cite{Friedman:1997yq}-\cite{Bouchard:2005ag}. In fact, it is simply part of the general analysis to show that a given bundle is somewhere slope-stable,
that is,  admits a connection obeying the Hermitian
Yang-Mills equations. Using our results, this structure provides information about which terms could possibly appear in the associated four-dimensional superpotential. Having
access to such information early in the construction of a model
can be extremely useful. Instead of first computing the details of a
compactification, calculating the Yukawa couplings and
discovering, for example, that the top quark mass vanishes, one can analyze the broad
features of the allowed interactions at the start to see if the model
has any possibility of being phenomenologically viable. 

Green-Schwarz anomalous $U(1)$ symmetries, and the phenomenological constraints arising from them, have been used extensively in model building in Type II theories (for example, see \cite{Ibanez:2001nd,Antoniadis:2002cs}) and have played an important role in recent work on D-brane instantons \cite{Ibanez:2006da}-\cite{Cvetic:2009ah}. In addition, such effects have been used to discuss Yukawa textures and hierarchies in F-theory \cite{Cecotti:2009zf,Marsano:2009wr}. However, it is important to note that the source of the anomalous $U(1)$ symmetries in the present work-- namely, their origin in the global stability structure of the K\"ahler cone-- is entirely new and provides an interesting contrast to the way that such symmetries arise in other contexts in string theory. It is also worth noting that the Yukawa textures explored in this work are distinct from those previously explored in the heterotic context \cite{Braun:2005xp,Braun:2006me}.

For specificity, the explicit examples in this paper involve bundles defined by the monad construction \cite{Anderson:2009mh}-\cite{yukawa} and by extension \cite{burt,AG,Braun:2005zv} over complete intersection Calabi-Yau threefolds \cite{Candelas:1987kf}. However, our results and conclusions are completely general and apply to any holomorphic vector bundle with K\"ahler cone sub-structure defined on any Calabi-Yau manifold.
The paper is structured as follows. In the next section,
we review general heterotic compactifications as well as the mathematics and associated effective field theories of stability walls.
In Section \ref{yukawawall}, we describe the Yukawa textures that can result
from the presence of the simplest kind of stability wall. Sections
\ref{onewall2D} and \ref{2walls} discuss two generalizations of this; first, to
stability walls with more complicated internal structure and, second,
to the case where multiple stability walls are present in a single
K\"ahler cone. A phenomenologically
interesting example of these ideas is presented in Section \ref{threegen}.
Constraints imposed by stability walls on massive vector-like pairs of
matter multiplets are analyzed in Section \ref{pairs}.  In Section
\ref{conclusions}, we give our conclusions. The paper has two
appendices. Appendix A presents a list of all possible Yukawa textures
that can result from the simplest kind of stability walls. In Appendix
B, we discuss some technical details associated with the
phenomenologically realistic example of Section \ref{threegen}.

\section{Heterotic Vacua and Vector Bundle Stability}
\label{section1}

\subsection{General Definitions}
\label{normal}

In $E_8 \times E_8$ heterotic string and
M-theory, compactification on a smooth Calabi-Yau threefold is not
sufficient to ensure that the four-dimensional effective theory is
${\cal N}=1$ supersymmetric.
Since heterotic compactifications
necessarily include background gauge fields, supersymmetry is also dependent on the choice of 
gauge connection and its properties.  Dimensional reduction yields the
well-known result that to preserve supersymmetry, these gauge
fields must solve the Hermitian Yang-Mills equations
\begin{eqnarray}
\label{hym}
g^{a \bar{b}} F_{a \bar{b}} =0 \;,\; F_{ab}=0 \;,\; F_{\bar{a}\bar{b}}=0~. 
\end{eqnarray}
The latter two equations simply require that the connection be
holomorphic.  However, the first condition, $g^{a \bar{b}} F_{a
  \bar{b}} =0$, is a notoriously difficult partial differential
equation to solve, involving not only the gauge connection but also
the Calabi-Yau metric - an object known only numerically at best
\cite{Douglas:2006hz}-\cite{Braun:2008jp}. Fortunately, the
Donaldson-Uhlenbeck-Yau theorem \cite{duy1,duy2} presents tractable
algebraic conditions under which a solution is guaranteed to exist,
without having to construct it explicitly.

The content of the Donaldson-Uhlenbeck-Yau theorem,
as relevant in this context, may be stated as follows: {\it On a
  compact K\"ahler manifold, a vector bundle $V$ admits a connection
  solving the Hermitian-Yang-Mills equations if and only if $V$ is a poly-stable holomorphic vector
  bundle of zero slope}.  To explain this statement, we must
describe what a poly-stable holomorphic vector bundle is and define
the notion of {\it slope}. The slope of a vector bundle (or sheaf) ${\cal{F}}$ is given by the integral
\begin{eqnarray}
\label{slope}
\mu(\cF) =\frac{1}{rk(\cal{F})} \int_X c_1(\cF) \wedge J \wedge J \ ,
\end{eqnarray} 
where $X$ is the Calabi-Yau manifold with K\"ahler form $J$ and $c_1(\cF)$ is the first Chern class of  
$\cF$. 
A vector bundle, $V$, is said to be {\it stable} for a given choice of
the K\"ahler form if every sub-bundle\footnote{Really a subsheaf. Stability of a vector bundle is defined so that $\mu(\cF) < \mu(V)$ for all torsion-free sub-sheaves, $\cF \in V$ with  $\textnormal{rk}(\cF) < \textnormal{rk}(V)$. However, for the examples in this work, all the de-stabilizing sub-objects will be bundles and hence for simplicity we will not discuss sheaves.} actually defined with respect to to $\cF$
in $V$ with $\textnormal{rk}(\cF) < \textnormal{rk}(V)$ has slope
strictly less than that of the bundle itself. That is, 
\beq\label{stab} 
\mu(\cF) <\mu(V)~~~~\forall~ \cF \ {\rm in}\ V~.  
\eeq A bundle is called {\it semi-stable} if
$\mu(\cF) \leq \mu(V)$ for all proper sub-bundles $\cF$.
We note that it is not stability that appears in the statement of the
Donaldson-Uhlenbeck-Yau theorem, but poly-stability. A bundle is
{\it poly-stable} if it is a direct sum of stable bundles, all of which have
the same slope. That is, 
\beq\label{poly_stab}
 V=\bigoplus_i V_i~~~
\mu(V)=\mu(V_i)~~~\forall i~.  
\eeq
Clearly, all poly-stable bundles are semi-stable, but
the converse does not hold. 
Hence, semi-stable bundles will be of interest to us only when they are
also poly-stable. 

An essential property, both mathematically and for physical
applications, of the notion of stability - as well as semi-stability
and poly-stability - is that it depends explicitly on the choice of
K\"ahler form $J$ on $X$. To understand the exact meaning of this, it
is useful to expand $J$ in a basis $J_{i}$, $i=1,\ldots h^{1,1}(X)$, of
$(1,1)$-forms as $J=t^i J_i$. The coefficients $t^{i}$ are the
K\"ahler moduli. Inserting this into \eqref{slope}, the slope of any
sub-bundle $\cal{F}$ can be written as
\begin{equation}
\mu ({\cal{F}}) = \frac{1}{rk({\cal{F}})} d_{ijk} c^{i}_{1} ({\cal{F}})t^{j}t^{k},
\label{burt1}
\end{equation}
where $d_{ijk}=\int_{X} {J_{i} \wedge J_{j} \wedge J_{k}}$ are the triple intersection numbers of $X$ and $c_{1}({\cal{F}})=c_{1}^{i}({\cal{F}})J_{i}$. That is, the slope of each sub-bundle ${\cal{F}}$ is a calculable function of the K\"ahler moduli $t^{i}$. It follows that whether or not a bundle is stable, poly-stable or semi-stable is, in general, a function of where one is in K\"ahler moduli space. {\it A vector bundle $V$  which is stable in one region of the K\"ahler cone of $X$ may not necessarily be stable in another}. 

\subsection{Stability Walls and K\"ahler Cone Substructure}
\label{wall}

How does one determine the the regions of stability/instability of a vector bundle? We begin by noting that the stability properties of a vector bundle for a choice of K\"ahler class\footnote{The choice of a K\"ahler form $J$, is referred to as a ``polarization'' in the mathematics literature.} $J$ will remain unchanged if that K\"ahler class is multiplied by a non-vanishing complex number. Hence, the stability properties of a bundle are the same along any one-dimensional ray in the K\"ahler cone. It follows that for a Calabi-Yau manifold with $h^{1,1}(X)=1$, a vector bundle will either be stable, or unstable, everywhere in the one-dimensional K\"ahler cone. We will, therefore, restrict our discussion to Calabi-Yau threefolds, $X$, with $h^{1,1}(X) \geq 2$. Now consider a holomorphic vector bundle, $V$, on $X$ such that for at least one choice of K\"ahler form - and, hence, for the ray it defines - the bundle is slope stable. In this paper, we take all slope-stable bundles to be indecomposable\footnote{Note that decomposable vector bundles $V=\bigoplus_i V_i$ can at best be best poly-stable, see \eref{poly_stab}.} with structure group $SU(n)$. Hence, the first Chern class satisfies $c_1(V)=0$. It follows from \eqref{slope} that the slope of $V$ also vanishes. Thus, for an $SU(n)$ bundle to be stable for a given value of the
K\"ahler moduli, the slope of each of its sub-bundles,
calculated using the corresponding K\"ahler form $J$, must be
negative. 

It is quite possible to find bundles for which the slopes of all sub-bundles remain negative everywhere in the K\"ahler cone (for example, the tangent bundle, $TX$, to the Calabi-Yau threefold). Any such vector bundle will admit an $SU(n)$ connection satisfying the Hermitian Yang-Mills equations for any values of the K\"ahler moduli. 
Now, however, consider a case where there is one particular sub-bundle $\cal{F}$ (itself stable) whose slope,
while negative for the polarization where the bundle $V$ is assumed stable, gets smaller and smaller as one moves in the K\"ahler cone, eventually going to zero. The condition $\mu ({\cal{F}})=0$ is one equation restricting $h^{1,1}$ K\"ahler moduli. That is, the vanishing of the slope of ${\cal{F}}$ defines a co-dimension one boundary - called a ``stability wall'' - in the K\"ahler cone. As we
cross this wall, this sub-bundle becomes positive in slope and
destabilizes the vector bundle. 
That is, the bundle no longer satisfies the Hermitian Yang-Mills equations and supersymmetry is broken.
For such bundles, the K\"ahler cone has sub-structure
\cite{Anderson:2009sw,Anderson:2009nt,Sharpe:1998zu}; that is, it can  split into separate ``chambers'' with respect to the stability properties of $V$. In
one of these chambers a solution to \eqref{hym} can be found, and in the
others it can not. This stability induced sub-structure, and the effective field theory \cite{Anderson:2009sw,Anderson:2009nt} description of it, will be of central importance to this work.

On the boundary between a supersymmetric and non-supersymmetric
chamber of the K\"ahler cone, we know from the proceeding discussion
that there is a sub-bundle ${\cal F}$ injecting into the bundle $V$ which
has the same slope as the bundle itself. That is, we can write an injective morphism $ 0 \to \cF \to V \to \ldots~$. Coherent sheaves form an
Abelian category and, thus, one may always write a cokernel, ${\cal K}=V/\cF$, to
form a short exact sequence and re-express the  bundle as the extension
\bea \label{ext} 
0 \to {\cal F} \to V \to {\cal K} \to 0 \ . 
 \eea 
In other
words, no matter how the bundle was originally defined, if it has a
stability wall then it may be written as an
extension\footnote{Strictly speaking, this is true if the bundle has a
stability wall caused by a single destabilizing sub-bundle. We will
discuss more general cases in later sections.}.

Given that on the stability wall 
${\cal F}$ injects into $V$ and has equal slope, the only way in which $V$ can preserve supersymmetry,
according to the Donaldson-Uhlenbeck-Yau theorem, is if it splits into
a direct sum of two pieces. In other words, supersymmetry is only
preserved on the wall when the sequence \eqref{ext} splits and 
\begin{equation}
V={\cal F} \oplus {\cal K} \ .
\label{burt4}
\end{equation}
Is this always possible? To answer this, note that the set of equivalent extensions, $V$, in \eqref{ext} is described by the group
$\textnormal{Ext}^1({\cal K},{\cal F})$. The split configuration, \eref{burt4},
corresponds precisely to the zero element in that space
\cite{Anderson:2009nt}.  Thus, as we approach a stability wall in
K\"ahler moduli space, the system can continue to preserve ${\cal
  N}=1$ supersymmetry\footnote{Mathematically, this statement can be understood by saying that on the wall, the semi-stable bundle, $V$, is an element of an S-equivalence class \cite{Huybrechts}. Since each S-equivalence class contains a unique poly-stable representative, it is always possible for the bundle to decompose as in \eref{burt4}.}. The price for this, however, is a decomposition
of the bundle into two pieces $V={\cal F} \oplus {\cal K}$. Such a
splitting of the bundle on the stability wall corresponds physically
to a change in the group in which the gauge field background of the
compactification is valued. If we begin with a stable $SU(n)$ bundle
$V$ then, at the stability wall, the structure group changes to
$S[U(n_1)\times U(n-n_1)]$ where $n_1$ is the rank of $\cF$. The exact
splitting depends on the the choice of structure group $SU(n)$ in the
stable chamber and exactly which sub-bundle destabilizes the bundle at
the stability wall. Generically, however, we can see that the effect
of the splitting \eqref{burt4} will be to change the low-energy
effective theory associated with this compactification. If we denote
the commutant within $E_8$ of $SU(n)$ as $H$ - the symmetry group of
the four-dimensional theory in the stable chamber - then the commutant
of $S[U(n_1)\times U(n-n_1)]$ will be enhanced by one additional
anomalous gauged $U(1)$ symmetry to $H\times U(1)$.

\subsection{Example: An $SU(3)$ Heterotic Compactification}\label{firsteg}

To give a concrete example of such a compactification, consider the
Calabi-Yau threefold defined by a bi-cubic polynomial in
$\mathbb{P}^2 \times \mathbb{P}^2$ \begin{eqnarray}
\label{eg1cy3}
\left[ \ba{c |c }
\mathbb{P}^2 & 3 \\
\mathbb{P}^2 & 3
\ea \right]^{2,83} \ ,
\end{eqnarray}
where the superscripts are $h^{1,1}$ and
$h^{1,2}$ respectively.
On this manifold, let us define a holomorphic vector
bundle, $V$, with structure group $SU(3)$. The bundle is given by a two-step process. First construct a rank $2$ bundle, ${\cal G}$, by the so-called monad construction \cite{Anderson:2009mh}- \cite{yukawa} via the short exact sequence
\begin{equation}
0 \to {\cal G} \to {\cal O}(1,0)^{\oplus 3} \oplus {\cal O}(1,1) \stackrel{f}{\to} {\cal O}(1,2) \oplus {\cal O}(2,2) \to 0 \ .
\label{eg1bundle}
\end{equation}
${\cal G}$ is defined in terms of the bundle morphism, $f$, above as ${\cal G}=ker(f)$. Next, we proceed to build the rank three bundle, $V$, out of the line bundle $\cO(-1,3)$ and ${\cal G}$, by ``extension''. That is, 
\begin{equation}
0 \to {\cal O}(-1,3) \to V \to {\cal G} \to 0 \ .
\label{eg1bundle 2}
\end{equation}
The manifold \eqref{eg1cy3} is a complete intersection
Calabi-Yau manifold \cite{Candelas:1987kf}. There are only
two independent harmonic $(1,1)$ forms on $X$, and a basis $J_1,J_2$ may be chosen which are the restrictions of the K\"ahler forms of each $\mathbb{P}^2$ to the Calabi-Yau hypersurface. In \eref{eg1bundle} and \eref{eg1bundle 2} above, $\cO(k,m)$ denotes a line bundle on $X$. The pair of integers $(k,m)$ fully specify the line bundle on $X$ by defining its first Chern class, $c_1(L)=kJ_1 +mJ_2$.  

The extension bundle $V$ in \eref{eg1bundle 2} can be
viewed as a non-trivial deformation of the direct sum of the two pieces, ${\cal G}$ and ${\cal O}(-1,3)$. At a generic point in its moduli space, that is, for generic choices of
the maps in \eqref{eg1bundle} and \eqref{eg1bundle 2}, the bundle $V$
has structure group $SU(3)$. For some choices of K\"ahler form, it is slope stable and, hence, 
$V$ corresponds to an $SU(3)$ valued solution to \eqref{hym}. To find where $V$ is stable, one must find all sub-bundles $\cal{F}$, calculate their slopes and check that these are all negative. For such an analysis, see, for example, \cite{Anderson:2009nt,stability_paper}. Here, we simply present our results. 

\begin{figure}[!ht]
  \centerline{\epsfxsize=5in\epsfbox{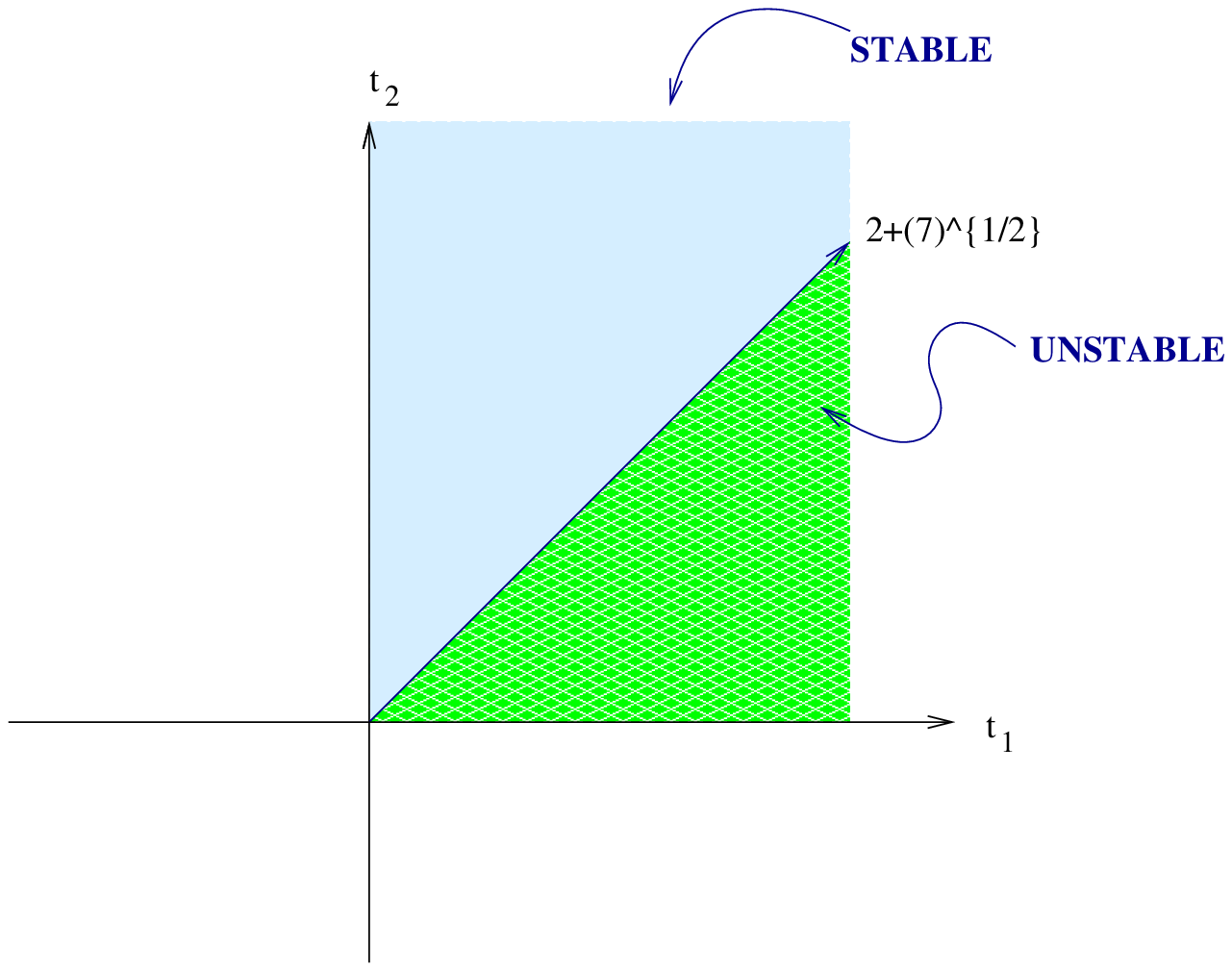}} \mycaption{
    The K\"ahler cone and regions of stability/instability for
    Calabi-Yau threefold \eqref{eg1cy3} and the bundle \eqref{eg1bundle 2}. The stability wall generated by $\cO(-1,3)$ in $V$ occurs on the line with slope $t^{2}/t^{1}=2+\sqrt{7}$.}
\label{figure1}
\end{figure}

Figure \ref{figure1} shows the two-dimensional K\"ahler cone of Calabi-Yau
threefold \eqref{eg1cy3}.  The physical K\"ahler cone, where the Calabi-Yau is
positive in volume and non-singular, is the complete
colored region. The light blue, upper region, in Figure \ref{figure1}
is the set of polarizations
for which the slope of each sub-bundle of the bundle is
negative and, hence, the bundle is stable.

Now note that the description of bundle \eqref{eg1bundle 2} 
is already in
the form \eqref{ext}.  We can, therefore, simply read off ${\cal F}$
and ${\cal K}$ from \eref{ext} as 
  \bea
&&{\cal F} = {\cal O}(-1,3) \label{egdec} \ ,\\
&& {\cal K}={\cal G}~~~\text{where}~~~ 0\to {\cal G} \to {\cal O}(1,0)^{\oplus 3} \oplus {\cal O}(1,1) \to
{\cal O}(1,2) \oplus {\cal O} (2,2) \to 0 \ . \label{egdec2} \eea
It follows that the bundle $V$ in \eref{eg1bundle 2} has a stability
wall of the kind we have been describing. This wall is shown as the 
line in Figure \ref{figure1}. It separates the region of stability of $V$ from its region of instability. The splitting $V \rightarrow \cF \oplus {\cal K}$ on the stability wall corresponds
physically to a change in the group in which the gauge field
background is valued. For this example, these
gauge fields change from being valued in $SU(3)$ in the
interior of the supersymmetric region, to being valued in $S[U(2)\times U(1)]\cong
SU(2) \times U(1)$ on the stability wall.  Recall that the four-dimensional symmetry group is the commutant of the structure group of the bundle
inside $E_8$. Thus, while the low-energy gauge group is simply $E_6$ in
the stable region, an extra Abelian factor appears when the
moduli are exactly on the stability wall. Here, the effective gauge symmetry is enhanced to
$E_6 \times U(1)$.

\subsection{The Particle Spectrum and Quantum Numbers}\label{first_spectra}

An analysis of the particle spectrum and the associated quantum numbers, both in the interior of the stable region of the K\"ahler cone as well as on a stability wall, is most easily presented in the context of an explicit example. Let us use the Calabi-Yau threefold, $X$, and the $SU(3)$ vector bundle presented in \eqref{eg1cy3} and \eqref{eg1bundle},\eqref{eg1bundle 2} above. 

In the interior of the stable region, the background gauge fields have structure group $SU(3)$ and the symmetry group of the the four-dimensional effective field theory is $E_6$.
Computing the matter spectrum of this low energy theory is an exercise in group theory and bundle
cohomology \cite{Green:1987mn}. All matter fields in the
ten-dimensional theory are valued in the ${\bf 248}$
representation of $E_8$. The matter multiplets that
appear in the four-dimensional spectrum are determined by the
branching of this representation under 
\begin{eqnarray}
\label{eg1grouptheory}
E_8 &\supset& E_6 \times SU(3) \\ 
{\bf 248} &=& ({\bf 78},{\bf 1}) + ({\bf 8},{\bf 1}) + ({\bf 27},{\bf 3}) + ({\bf \overline{27}},{\bf \overline{3}}) \  .
\end{eqnarray}
The first number in the brackets above is the dimension of a
representation of $E_6$ and the second the dimension of a
representation of $SU(3)$.
To find the multiplicity of each term, one must compute 
the number of zero-modes of the associated twisted Dirac operators on the internal space \cite{Green:1987mn}. 
This is given by the dimension of the relevant bundle-valued
cohomology group. 
The group representations, four-dimensional field names  and the associated cohomologies for a {\it generic} $E_6$ theory are indicated in the first three columns of
Table \ref{table1}.
\begin{table}[t]
\begin{center}
\begin{tabular}{|c|c|c||c|}
  \hline
  Representation & Field Name & Cohomology & Multiplicity \\ \hline
  $({\bf 8},{\bf 1})$ &  $\phi$ & $h^1(X,V \otimes V^*)$ & 87 \\ \hline
  $({\bf 27},{\bf 3})$ & $F^I$ &  $h^1(X,V)$ & 39 \\ \hline
  $({\bf \overline{27}},{\bf \overline{3}})$ & $\overline{F}^A$ &$h^1(X,V^*)$ &0
  \\ \hline
\end{tabular}
\mycaption{The representations, field content and the associated cohomologies for a {\it generic} $E_6$ theory. For a Calabi-Yau threefold $h^1(X,\cO_X)=0$ and $n_{78}=h^0(X,\cO_X)=1$.
The multiplicities for the {\it specific} indecomposable rank $3$ vector bundle $V$ defined in \eref{eg1bundle 2} are given in the fourth column. }
\label{table1}
\end{center}
\end{table}

The dimensions of the cohomologies for the {\it specific} bundle $V$ in example \eref{eg1bundle 2}
are presented in the fourth column. We see, in particular, that 
we have $39$ ${\bf 27}$ dimensional
representations of $E_6$. At this stage, there is nothing to suggest
any sort of ``texture'' in the cubic self-interactions of these
fields. Generically, one would expect all Yukawa terms which are allowed by $E_6$ gauge symmetry to appear.
In fact, this is {\it not} the
case, as we will show in the next section.

For a K\"ahler form {\it on the stability wall}, the background gauge fields are valued in 
$S[U(2)\times U(1)] \cong SU(2) \times U(1)$ and
the symmetry group of the the four-dimensional theory is $E_6 \times U(1)$.
The method for computing the spectrum and quantum numbers on the stability wall is analogous to the procedure above. The only
difference is that one now takes the gauge bundle to be $V={\cal F} \oplus
{\cal K}$, rather than indecomposable and rank 3. 
The group theory which determines which multiplets
can appear in four dimensions is now   \bea
\label{eg1splitgrouptheory}
E_8 &\supset& E_6 \times SU(2) \times U(1) \\
{\bf 248} &=& ({\bf 1},{\bf 1})_0 + ({\bf 1},{\bf 2})_3 + ({\bf
  1},{\bf 2})_{-3} + ({\bf 1},{\bf 3})_0 + ({\bf 78},{\bf 1})_0 \\
\nonumber && + ({\bf 27},{\bf 1})_{-2} + ({\bf 27}, {\bf 2})_1 + ({\bf
  \overline{27}}, {\bf 1})_2 + ({\bf \overline{27}},{\bf 2})_{-1} \ . \eea
Note that each multiplet has an additional quantum number associated with the $U(1)$ factor in the effective theory. The group representations, four-dimensional field names  and the associated cohomologies for a {\it generic} $E_6 \times U(1)$ theory are indicated in the first three columns of
Table \ref{table2}. The multiplicity is 
found by calculating the dimension of each cohomology.
The results for the decomposition $\cF \oplus {\cal K}$ associated  with the {\it explicit} example 
\eref{egdec},\eref{egdec2} are given in the fourth column. 
\begin{table}[t]
\begin{center}
\begin{tabular}{|c|c|c||c|}
  \hline
  Representation & Field Name  & Cohomology & Multiplicity \\ \hline
   $({\bf 1},{\bf 2})_3$ &  $C_1^p$ & $h^1(X, {\cal F}^*\otimes {\cal K})$ &0 \\ \hline
  $({\bf 1},{\bf 2})_{-3}$ & $C_2^Q$ &  $h^1(X,{\cal F}\otimes {\cal K}^*)$ &21 \\ \hline
  $({\bf 1},{\bf 3})_0$ &  $\psi$ & $h^1(X,{\cal K} \otimes {\cal K}^*)$ &67 \\ \hline
  $({\bf 27},{\bf 1})_{-2}$ &  $F_1^i$ & $h^1(X,{\cal F})$ & 3 \\ \hline
  $({\bf 27},{\bf 2})_1$ &  $F_2^{\eta}$ & $h^1(X,{\cal K})$ & 36\\ \hline
  $({\bf \overline{27}},{\bf \overline{1}})_2$ & $\overline{F}_1^a$ & $h^1(X,{\cal F}^*)$ & 0
  \\ \hline
  $({\bf \overline{27}},{\bf \overline{2}})_{-1}$ &   $\overline{F}_2^{\alpha}$ & $h^1(X,{\cal K}^*)$ &0
  \\ \hline
\end{tabular}
\mycaption{The representations, field content and the cohomologies of a {\it generic} $E_6\times U(1)$ theory associated with a poly-stable bundle $\cF \oplus {\cal K}$ on the stability wall. 
Note that $h^1(X,{\cal F}\otimes {\cal F}^*)$ vanishes here since $\cal{F}$ is a line bundle.
The multiplicities for the {\it explicit} bundle defined by \eref{burt4} and \eref{egdec},\eref{egdec2} are shown in the fourth column.} \label{table2}
\end{center}
\end{table}
It is important to note here that the extra $U(1)$ symmetry is
Green-Schwarz anomalous, as described in detail in
\cite{Anderson:2009sw,Anderson:2009nt}. Thus, the usual anomaly cancellation constraints on
the charges do
not apply. For the general form of $U(1)$ charges possible in the
present context, see \cite{Lukas:1999nh}. We will come back to the
anomalous nature of this $U(1)$ in the following sections.

One obvious question is: what is the relationship between the particle spectrum on the stability wall, given in Table \ref{table2}, and the manifestly different spectrum in the interior of the stability region, presented in Table \ref{table1}? Furthermore, how does one relate their two four-dimensional field theories?
To answer these questions, we construct the effective theory on the stability wall and then consider small perturbations into the interior of the slope-stable region.

\subsection{Connecting the Two Theories} \label{connecting}

The effective theories associated with the stable bundle $V$ and the poly-stable bundle $\cF \oplus {\cal K}$, described generically in Section \ref{wall}, can be related by considering the vacuum near the stability wall. This relationship is most easily illustrated using the specific example in Subsection \ref{firsteg}. Begin with the K\"ahler moduli of the $E_6 \times U(1)$ theory on the stability wall in Figure \ref{figure1}.  Then vary them continuously, moving  away from the boundary and into the stable region of the K\"ahler cone. This should reproduce the physics of the $E_6$ compactification.

As shown in  \cite{Anderson:2009sw,Anderson:2009nt}, the effective theory both on and near the stability wall is described by a D-term
associated with the enhanced gauged $U(1)$ factor. It is given by 
\bea \label{dterm} 
D^{U(1)} =
\frac{3}{16} \frac{\epsilon_S \epsilon_R^2}{\kappa_4^2}
\frac{\mu({\cal F})}{ {\cal V}}  -
\frac{1}{2}\sum_{P, \overline{Q}} Q_{2}G_{P \overline{Q}} C_2^P
\overline{C}_2^{\overline{Q}} \  ,
\eea
where the charge\footnote{Note that locally $S[U(2)\times U(1)]\approx SU(2)\times U(1)$. Globally, however, there is different overall normalization on the $U(1)$ which commutes with this group within $E_8$. In Ref. \cite{Anderson:2009nt} the $U(1)$ normalization was chosen consistent with the global description. In this work, since we are interested only in gauge invariant quantities (where overall normalization does not matter) we have chosen for simplicity the {\it local} charge normalizations consistent with Ref. \cite{Slansky:1981yr}. } $Q_{2}=-3$.
The first term is a K\"ahler modulus dependent ``Fayet-Iliopoulos'' (FI) term.
This is a multiple of the slope of the destabilizing
sub-bundle, divided by the volume ${\cal V}$ of
the Calabi-Yau threefold. The constants $\epsilon_S$ and $\epsilon_R$ are the usual expansion
parameters defining four-dimensional heterotic M-theory \cite{Lukas:1998hk}. 
It follows from the discussion in Subsection \ref{firsteg} that the FI term is positive in the non-supersymmetric
(dark shaded) region of Figure \ref{figure1}, negative in the stable (light shaded) region, and
vanishes on the boundary line between the two. The second term is the usual
contribution to a D-term from charged matter.  The fields shown in \eref{dterm} are the $E_6$ 
singlets $C_2$ in Table \ref{table2}. 
The positive definite field space metric
$G_{P \overline{Q}}$ appears since they are not generically canonically
normalized. In the explicit example of Subsection \ref{firsteg}, there are no $C_{1}$ fields in the spectrum. The $27$ and $\overline{27}$ representations in Table \ref{table2}, which are
also charged under $U(1)$, should appear in this D-term as well. However, 
the $E_6$ D-terms force the vevs of these fields to vanish, and hence, they can be safely ignored in the following discussion.

Using the D-term \eqref{dterm}, one can concretely specify the relationship between the effective theories in the stable, poly-stable and unstable regions of Figure \ref{figure1}. On the stability wall, $\mu({\cal{F}})=0$ and the $D^{U(1)}$ contribution to the potential is minimized for $\left<C^{P}_{2}\right>=0$. Hence, the theory is $E_{6} \times U(1)$ invariant with the spectrum of massless fields given in Table \ref{table2}. Strictly speaking, the $U(1)$ factor is Green-Schwarz anomalous. Hence, the associated gauge boson is
not massless, even on the stability wall. The mass of this
gauge boson was computed in
\cite{Anderson:2009sw,Anderson:2009nt,Lukas:1999nh,Blumenhagen:2005ga}. On the stability wall it was found to be  \bea
\label{u1masswall} 
m_{U(1)}^2=\frac{1}{s}\left(\frac{(3 \epsilon_S
    \epsilon_R^2)^2}{256\kappa_4^2} c_1^i(\cF)c_1^j(\cF)G_{ij}
  \right) \ ,\eea 
  where $G_{ij}=-\frac{\partial^2 \text{ln}{\cal V}}{\partial t^{i} \partial t^{j}}$ 
and $s=ReS$ is the real part of the dilaton. This is parametrically lighter than the compactification scale
and, hence, the Abelian gauge boson must be included in the four-dimensional effective theory.  
What happens to D-term \eqref{dterm} as one moves continuously off the stability wall and into the stable region of moduli space? 
Here, $\mu(\cF) <0$ and the $C_2$ fields acquire non-zero vevs so as to set $D^{U(1)} = 0$ and minimize the potential. The $\left<C^{P}_2\right> \neq 0$ vevs thus spontaneously  break
$U(1)$, reducing the symmetry to a pure $E_6$ gauge
theory. 
Specifically, 
the mass of the $U(1)$ gauge boson is enhanced \cite{Anderson:2009sw,Anderson:2009nt} from \eqref{u1masswall} to 
\bea
\label{u1mass} 
m_{U(1)}^2=\frac{1}{s}\left(\frac{(3 \epsilon_S
    \epsilon_R^2)^2}{256\kappa_4^2} c_1^i(\cF)c_1^j(\cF)G_{ij}
  +\frac{9}{4} \sum_{P,\bar{Q}} G_{P \bar{Q}} \langle C_2 \rangle^P
  \langle \bar{C}_2 \rangle^{\bar{Q}}\right) \ .\eea 
As $\left<C_2 \right>$ increases in magnitude, their contribution
drives the $U(1)$ gauge boson mass above the compactification
scale. It must then be integrated out and removed
from the four-dimensional theory.  This process, both on and off of the stability wall, is simply a Higgs
effect. 

Expanding each field as a small fluctuation around its vev and
using $\left<D^{U(1)}\right>=0$, the D-term \eqref{dterm} is given to
linear order by
\bea
\label{linearcombo} \delta D^{U(1)} = -\frac{3}{16} \frac{\epsilon_S
  \epsilon_R^2}{\kappa_4^2} G_{jk} c_1^j({\cal F}) \delta t^k -
\frac{3}{2} \sum_{P,\bar{Q}} G_{P \bar{Q}} \left( \left< C_2^P\right>
  \delta \bar{C}_2^{\bar{Q}} + \delta C_2^P \left< \bar{C}_2^{\bar{Q}}
  \right> \right) \ . \eea 
From $V=\frac{1}{2s} (\delta D^{U(1)})^{2}$, canonically normalizing the kinetic energy and using \eqref{u1mass}, one can extract that massive Higgs field as a linear combination of $\delta t^{k}$ and $\delta C_{2}^{P}$ fluctuations. This is explicitly discussed in 
\cite{Anderson:2009sw,Anderson:2009nt}. Suffice it here to say that on the stability wall, the Higgs 
field reduces to the linear combination of $\delta t^{1}$, $\delta t^{2}$ perpendicular to the line in Figure \ref{figure1}.  The associated linear combination of K\"ahler moduli axions acts as the Goldstone boson and is ``eaten'' so as to give additional mass to the $U(1)$  gauge boson. Thus, near the stability wall one entire complex linear combination of K\"ahler moduli becomes heavy due to the Higgs mechanism. As one moves away from the stability wall in K\"ahler moduli space, the vevs of the $C_2$ fields adjust so as to minimize the potential. As discussed in \cite{Anderson:2009sw,Anderson:2009nt},
the $\delta C_{2}^{P}$ terms quickly become the dominant contribution to the Higgs field. 
Thus, in the stable region far from the wall, essentially one complex
$C_2$ field is lost to the Higgs effect. 

One can now explicitly describe the transition from the massless $E_{6} \times U(1)$ spectrum on the stability wall, given in Table \ref{table2}, to the $E_{6}$ zero-mode spectrum in the interior of the stable region, Table \ref{table1}.
Of the $21$ $C_2$ fields on the stability wall, $1$ of them is lost
through the Higgs mechanism as one moves into the stable
region. Integrating out the heavy $U(1)$ gauge boson, the -$3$ charge
of the remaining $20$ $C_2$ fields can be ignored. These combine with
the $67$ $\psi$ fields of Table \ref{table2} to correctly reproduce
the $87$ uncharged bundle moduli of the $E_{6}$ theory in
Table~\ref{table1}.  Furthermore, when the $U(1)$ symmetry is
integrated out, the quantum numbers distinguishing the two types of
${\bf 27}$ fields at the stability wall, $3$ with $U(1)$ charge -$2$
and $36$ with charge $+1$, no longer label the spectrum. Thus, we find
the expected $39$ ${\bf 27}$ fields of Table \ref{table1}. This
correspondence between the massless spectrum near a stability wall and
that in the interior of the stable region was proven in complete
generality in \cite{Anderson:2009nt} \footnote{Note that in moving
  between the stable region and the poly-stable wall, only the chiral
  asymmetry need be preserved.
The actual number of $27$ and $\overline{27}$ representations does not necessarily remain the same. In particular, massless vector-like pairs on the stability wall
can become massive in the stable region of moduli space. 
We will return to the issue of massive vector-like pairs, and possible constraints on them, in Section \ref{pairs}.}.

Finally, let us start once again on the stability wall. Now, continuously vary the moduli into the unstable region, where $\mu({\cal{F}}) > 0$. In principle, the $C_{1}$ fields with $U(1)$ charge $Q_{1}=+3$ could cancel the positive FI-term. However, for the bundle in \eref{egdec} and \eref{egdec2}, we see from Table \ref{table2} that there are no $C_1$ fields present in the spectrum. Therefore $D^{U(1)} \neq 0$ and supersymmetry is broken, as we expect from the stability analysis.

\subsection{The Charged Bundle Moduli $C_i$ and Branch Structure}\label{csection}

In this subsection, for specificity, we consider rank three bundles
whose K\"ahler cone contains at least one stability wall. Furthermore,
our analysis is confined to a single wall where the $SU(3)$ structure
group decomposes into $S[U(2) \times U(1)]$. Hence, the $E_{6}$ gauge
group is enhanced by a single $U(1)$ factor, giving rise to one
Abelian D-term in the effective theory. Our discussion will,
therefore, be applicable to the specific example discussed in
Subsections 2.3, 2.4 and 2.5, but will be considerably more
general. We emphasize that the type of conclusions drawn from this
analysis will remain unchanged for bundles of higher rank, and for
stability walls described by more than one D-term.

Consider a general rank three bundle $V$, destabilized by a single sub-bundle $\cF$ as in \eref{ext}, which generates K\"ahler cone sub-structure of the form discussed in Section \ref{wall}. In general, for such a bundle, there are precisely two types of bundle moduli {\it charged under the extended $U(1)$ symmetry}.
These are denoted $C_{1}$,$C_{2}$ and arise from the cohomologies shown in Table \ref{table2}. These charged bundle moduli, by acquiring vevs to cancel the FI-term, play a central role in controlling the supersymmetry of the theory. In the specific example of Section \ref{firsteg}, only negatively charged $C_{2}$ fields appeared in the spectrum. The D-term potential generated by these fields
exactly reproduced the regions of slope stability and instability shown in Figure \ref{figure1}.
For a more general bundle, however, it is possible that {\it both} fields $C_{1},C_{2}$ in Table \ref{table2} are present in the spectrum. In this case, the $U(1)$ D-term takes the form

\bea \label{dterm2} 
D^{U(1)} =
\frac{3}{16} \frac{\epsilon_S \epsilon_R^2}{\kappa_4^2}
\frac{\mu({\cal F})}{ {\cal V}} - \frac{3}{2} \sum_{p, \overline{q}}
G_{p \overline{q}} C_1^p \overline{C}_1^{\overline{q}} + \frac{3}{2}
\sum_{P, \overline{Q}} G_{P \overline{Q}} C_2^P
\overline{C}_2^{\overline{Q}} \ .
\eea 
Now there are {\it two} terms available to cancel the K\"ahler moduli dependent FI-term. As we will see however, they play very different roles and {\it $C_1,C_2$ can never obtain non-zero vevs simultaneously}. 

To show this, first note, that in addition to the D-term \eqref{dterm2}, one must also consider the superpotential. Again ignoring $E_6$ non-singlets, this can be written as \footnote{Our argument will be unchanged if we include in $W$ all higher powers of $C_1C_2$; that is, if $W \sim \sum_{n} (C_1C_2)^n$. Hence, we consider only the lowest order term.}
\bea
W = \lambda_0 (C_1 C_2) ^2 
\eea
where the indices on both fields and
couplings are suppressed. In the stable region of K\"ahler moduli space, the four-dimensional effective theories we are considering have supersymmetric, Minkowski
vacua. Therefore, as we vary the K\"ahler
moduli away from the stability wall into the $\mu(\cF)<0$ region of bundle $V$, we must preserve supersymmetry {\it and} avoid introducing a cosmological constant. The relevant equations, in
addition to the vanishing of D-term \eqref{dterm2}, are
\bea \nonumber
\partial_{C_1} W &=& \lambda_{0} C_2 (C_1 C_2) =0 \ , \\  \label{Weqns}
\partial_{C_2} W &=& \lambda_{0} C_1 (C_1 C_2) =0 \ , \\ \nonumber W &=&
\lambda_{0} C_1 C_2 C_1 C_2 = 0 \ .  \eea 
With $\mu(\cF)<0$ in \eref{dterm2}, one might suppose that to preserve supersymmetry, the fields $C_1$ and
$C_2$ could {\it both} get vevs such that the last two terms in $D^{U(1)}$ cancel the FI
term. 
However, substituting these two non-zero vevs into equations \eqref{Weqns}, it is clear that no such solution is possible. This is most easily verified by noting that, without loss of generality, one can choose a basis of field space so that only one of the $C_1$ fields and one of the $C_2$ fields has a non-vanishing vev. Thus, to move into the stable region of $V$ and obtain a
Minkowski vacuum, the only choice available is to take all $\left<C_{1}^{p}\right>=0$ and to choose non-vanishing $C_{2}^{P}$ vevs so that the first and last terms
in \eqref{dterm2} cancel. What happens in the chamber of $V$ where $\mu({\cal{F}})>0$? Here, it would {\it appear} from \eqref{dterm2}, \eqref{Weqns} that supersymmetry could still be preserved by the reverse happening; that is, $C_{1}$ fields getting non-vanishing vevs while all $C_{2}$ vevs are zero. However, as we show in the remainder of this subsection, 
within the context of our chosen geometry, i.e. the bundle $V$ defined by \eqref{ext}, {\it only $C_{2}$ fields can have non-zero vevs}. Hence, in the $\mu({\cal{F}})>0$ chamber of the K\"ahler cone supersymmetry is spontaneously broken by the D-term.

The key to explaining this fact, and distinguishing the fields
$C_1$ and $C_2$, can be found in the associated algebraic
geometry. 
Although they behave as charged matter fields on the stability wall,
the $C_{1}$,$C_{2}$ fields can also be viewed geometrically as the
moduli which control the ``mixing'' of the components of $\cF \oplus
{\cal K}$ together to form an indecomposable bundle.  To see this,
recall how matter fields arise in a heterotic compactification. For
dimensional reduction, the ten-dimensional $E_{8}$ gauge fields,
${\cal{A}}$, on the ``visible sector'' fixed plane are expanded in a
decomposition which is related to the bundle structure group. On the
stability wall, the relevant ansatz is \bea \label{reduction}
{\cal{A}}_b = A_b + C_1^{p } \omega_{p\;b}^{(1) x} T^{(1)}_{x} +
C_2^{P} \omega_{P\;b}^{(2) y} T^{(2)}_{y} + \ldots \eea Fields $A_{b}$
are the gauge connection valued in $S[U(2) \times U(1)]$. The dots
indicate terms involving other fields, such as $F$, from Table
\ref{table2}. From \eqref{eg1splitgrouptheory} we see that the adjoint
of $E_8$ breaks up into a series of pieces, one which is $({\bf
  1},{\bf 2})_3$ and another $({\bf 1},{\bf 2})_{-3}$, under the
branching to $E_6 \times SU(2) \times U(1)$. $T^{(1)}$ and $T^{(2)}$
in \eqref{reduction} are precisely these gauge group generators, with
the indices $x$ and $y$ running over the ${\bf 2}$ representation of
$SU(2)$. The symbols $\omega^{(1)}$ and $\omega^{(2)}$ denote harmonic
one-forms valued in ${\cal F}^* \otimes {\cal K}$ and ${\cal F}
\otimes {\cal K}^*$ respectively.  Hence, the number of $C_2$ fields
is found by counting the independent one-forms valued in ${\cal F}
\otimes {\cal K}^*$, while the $C_1$ fields arise as the independent
one-forms valued in ${\cal F}^* \otimes {\cal K}$.
This can be re-expressed in terms of 
Ext-groups \cite{AG} as  
\bea \label{mrext}
\textnormal{Number of}\;\; C_2\textnormal{'s} &=& h^1(X,{\cal F} \otimes
{\cal K}^*) = \textnormal{dim}(\textnormal{Ext}^1({\cal K},{\cal F})) \ , \\
 \textnormal{Number of}\;\;C_1\textnormal{'s} &=& h^1(X,{\cal F}^* \otimes {\cal K}) =
\textnormal{dim}(\textnormal{Ext}^1({\cal F},{\cal K})) \ . \nonumber
\eea
From \eqref{reduction} we see that, when we give a $C$ field a vev,
the ten-dimensional gauge connection changes its expectation
value. Equation \eqref{mrext} tells us what this change means
in terms of bundle structure. The Ext-groups correspond to the moduli spaces of {\it two different extension bundles} \cite{AG,burt},
\begin{eqnarray}
&& 0 \to \cF \to V \to {\cal K} \to 0~~~\leftrightarrow~~~ \textnormal{Ext}^1({\cal K},{\cal F}) \ , \label{branch_bundles} \\
&& 0 \to {\cal K} \to {\tilde V} \to \cF \to 0~~~\leftrightarrow~~~ \textnormal{Ext}^1({\cal F},{\cal K})  
 \label{branch_bundles2}
 \end{eqnarray}
respectively.
$V$ and
$\tilde{V}$ are referred to as an extension and its ``dual'' extension.
They are {\it not in general isomorphic}.

It follows from \eqref{mrext}, \eqref{branch_bundles} that when $\left<C_2\right> \neq 0$, $A$ in
\eqref{reduction} becomes an {\it irreducible} connection on V.
Similarly, comparing \eqref{reduction},
\eqref{mrext} and \eqref{branch_bundles2}, we see that giving $C_1$ a
vev corresponds to $A$ becoming an irreducible connection on ${\tilde{V}}$. 
However, since $V$ and $\tilde{V}$ are not isomorphic, for a given geometry, one can have {\it either} non-vanishing $C_{2}$ {\it or} non-vanishing $C_{1}$, but {\it not both}. 
This is the higher-dimensional manifestation of the statement derived in
effective field theory earlier in this subsection: $\left<C_1\right>$ and
$\left<C_2\right>$ can be never be non-zero simultaneously. Note that the bundle $V$ discussed at the beginning of this subsection is of the type  \eqref{branch_bundles}. This explains why only its $C_{2}$ fields can get a non-zero vev. Importantly, however, one could just as easily have analyzed the stability regions of
$\tilde{V}$ defined by \eqref{branch_bundles2}, where $C_{1}$ can be non-zero.  
These two {\it ``branches''} of the vacuum space, where $\left<C_2\right> \geq 0$, and $\left<C_1\right> \geq 0$, respectively, intersect at exactly one locus, the stability wall, where both vevs vanish and the
connection in $A$ lives on the bundle ${\cal F} \oplus {\cal K}$. Thus,
by changing the vevs of the four-dimensional fields, one can move
smoothly between non-isomorphic internal gauge bundles for 
heterotic compactifications\footnote{Note that the D-term in \eref{dterm2}, and its associated quantities, are only defined up to an overall sign. However, the relative sign between the $C$ terms and the FI term in \eref{dterm2} is fixed, and arises from the choice of embedding of $S[U(2)\times U(1)]$ (associated with $\cF\oplus {\cal K}$) inside $SU(3)$ (associated with $V$) \cite{Anderson:2009nt} and anomaly cancellation \cite{Lukas:1999nh}. Since we began by describing the geometry of the bundle in \eref{ext} and \eref{branch_bundles}, here we have chosen the sign conventions in \eref{dterm2} so that the FI term is equal to a positive multiple of $\mu(\cF)$. Had we begun with \eref{branch_bundles2} instead, the opposite sign convention could, of course, be taken. Note that since $\mu(\cF)=-\mu({\cal K})$, whichever sign convention is chosen, the sign of the FI term will be opposite in the two branches.}. In the following, we will 
discuss the
theory {\it corresponding to only one branch at a time}. 
A more detailed study of this stability wall induced
branch structure, and transitions between such theories, will appear
separately \cite{bundletransitions}. 

\section{Wall Induced Yukawa Textures}\label{yukawawall}

We can now turn to the main question of this
paper - can the existence of a stability wall constrain the physics of a compactification, even when the vacuum is in the interior of the stable region? The answer, as we will see, is affirmative. In this section, we continue to illustrate the main ideas using rank three bundles whose K\"ahler cone contains a stability wall where the $SU(3)$ structure group decomposes into $S[U(2) \times U(1)]$. Thus, on and near this wall, the $E_{6}$ gauge group is enhanced by a single $U(1)$ factor, giving rise to one Abelian D-term in the effective theory.  The types of conclusion drawn from this analysis remain unchanged for bundles of higher rank, and for stability walls with more than one D-term.

\subsection{Textures Near a Stability Wall}\label{textnearwall}

Consider a heterotic compactification associated with a bundle $V$ of
the form \eref{ext}.  {\it On and near the stability wall}, the
superpotential is constrained by the gauge symmetry of the
four-dimensional theory, {\it including the extra $U(1)$}. Using Table
\ref{table2}, the relevant matter field superpotential consistent with
gauge invariance is given by \footnote{Note that if some of the
  fields, such as the $C_1$'s, do not appear in the low energy
  spectrum, that is, if the cohomology $H^1({\cal F}^* \otimes {\cal
    K})$ in Table \ref{table2} vanishes as in the example of
  Subsection \ref{firsteg}, then the following discussion will lead to
  even more restrictive Yukawa textures.}  \bea \label{Wgeneral} W =
\lambda_0 (C_1 C_2)^2 + \lambda_1 F_1^3 C_1^2 + \overline{\lambda}_1
\overline{F}_1^3 C_2^2 + \lambda_2 F_1^2 F_2 C_1 +\overline{\lambda}_2
\overline{F}_1^2 \overline{F}_2 C_2 \\ \nonumber +\lambda_3 F_1 F_2^2
+ \overline{\lambda}_3 \overline{F}_1 \overline{F}_2^2 +\lambda_4
F_2^3 C_2 + \overline{\lambda}_4 \overline{F}_2^3 C_1 \ .  \eea Note that no quadratic
terms appear, since all of these superfields are zero-modes of the
compactification. Furthermore, terms of dimension six or higher in
$E_6$ non-singlet fields are not of interest to us, so we ignore
them. Finally, we have displayed only the lowest dimensional terms
required in our analysis. Each term can be multiplied by any positive
integer power of $C_1 C_2$. Such terms do not change the subsequent
analysis and, hence, in the interests of brevity, we suppress them.

{\it On the stability wall} $\mu({\cal F})=0$ and, hence, the FI term in
\eqref{dterm2} vanishes. In order to have both $D^{\textnormal{U(1)}}=0$
and a solution to \eqref{Weqns}, it follows that $\left< C_1
\right>=\left< C_2 \right>=0$. Substituting this into
\eqref{Wgeneral}, the most general tri-linear
couplings possible between $E_6$ families on the stability wall are  \bea
\label{Wwall} W^{\textnormal{wall}}_{\textnormal{Yukawa}} = \lambda_3
F_1 F_2^2 + \overline{\lambda}_3 \overline{F}_1 \overline{F}_2^2 \ . \eea
Note that only one type of coupling appears. All others, such as $F_1^3$, vanish. This is an {\it extremely} restrictive texture of Yukawa couplings. 
The fact that a Yukawa texture emerges precisely on the stability wall
is, perhaps, of limited interest. Although some model building has
been carried out on such a locus
\cite{Blumenhagen:2006ux}, it is more common to build standard model-like physics in the interior stable region. 
Let us analyze, therefore, what happens to 
the texture, \eqref{Wwall}, as we move into this chamber.

Consider a point in the stable region {\it close to, but not on}, the stability wall. 
Here $\mu({\cal F})<0$, which implies, through the vanishing of the D-term \eqref{dterm2} and equations \eqref{Weqns}, that $\left<C_1 \right>=0$ and
$\left< C_2\right> \neq 0$.  Using this in \eqref{Wgeneral},
the allowed cubic matter couplings become
\bea \label{Wnearwall}
W^{\textnormal{near wall}}_{\textnormal{Yukawa}} = \lambda_3 F_1
F_2^2 + \overline{\lambda}_3
\overline{F}_1 \overline{F}^2_2
+{\bf  \overline{\lambda}_1
\left< C_2^2 \right> \overline{F}^3_1 + \overline{\lambda}_2 \left<
  C_2 \right> \overline{F}^2_1 \overline{F}_2 + \lambda_4 \left< C_2 \right> F_2^3 } \ . \eea
Note that the non-zero $C_2$ vevs have allowed some Yukawa couplings missing in \eqref{Wwall} to ``grow back'' from higher dimensional terms. These are expressed in boldface. This is not true of all Yukawa couplings however. Specifically, the $F_1^3$ and ${\bar{F}}_2^3$ terms are still forbidden, despite the fact that the extended $U(1)$ gauge symmetry is spontaneously broken. That is, there remains a non-trivial texture.

Thus, we have demonstrated the existence of non-trivial Yukawa texture induced by
a stability wall, even for {\it small} deformations of the moduli into the stable region. 
However, can one extend the analysis of this subsection to moduli deep in the interior of the stable chamber? To answer this, let us recall the effective field theory descriptions associated with 1) being {\it on} the wall, 2) {\it near} the wall and 3) {\it far} from the wall in the stable region. {\it On the wall}, 
$\left<C_{1}\right>=\left<C_{2}\right>=0$, both $C_{1},C_{2}$ are massless and the $U(1)$ vector boson has a non-zero mass given in \eqref{u1masswall}. Since this mass is significantly smaller than the compactification scale, the extended $U(1)$ should {\it not} be integrated out of the low energy theory. The superpotential is then restricted by the $U(1)$ charges to expression \eqref{Wgeneral} and the Yukawa couplings to \eqref{Wwall}. Moving away from the wall, $\left<C_{1}\right>=0$ and $\left<C_{2}\right>\neq 0$.  
However, the non-zero vevs of the $C_{2}$ fields enlarge the mass of
the $U(1)$ gauge boson via expression \eqref{u1mass}, give an
equivalent mass to a linear combination of $\delta t^{k}$,$\delta
C_{2}$ and mass to one combination of $C_{1}$ fields. As long as the
mass of the $U(1)$ gauge boson remains controllably below the
compactification scale, the $U(1)$ should still {\it not} be integrated out of the
theory and the superpotential continues to be given by
\eqref{Wgeneral} and the Yukawa couplings by \eqref{Wnearwall}. This
defines what it means to be {\it near the wall}. What happens {\it far
  from the wall}? By definition, this occurs when $\left<C_{2}\right>$
approaches a value such that the two terms in \eqref{u1mass} become of
equal size. It then follows from \eqref{u1mass} that the $U(1)$ gauge
boson and the $\delta C_{2}$ masses, as well as the $C_{1}$ mass,
become as large as the compactification scale and, hence, these field
must be integrated out of the effective theory. There are two
consequences of this. First, the linear combination of $C_{2}$ fields
with the non-zero vev is no longer in the spectrum and, hence, one can
not write higher dimension terms proportional to powers of
$\left<C_{2}\right>$ as in \eqref{Wnearwall}. Second,
the ${\bf 27}^{3}$ that do occur are no longer {\it necessarily}
constrained by the $U(1)$ quantum numbers. Hence, it would appear that
the Yukawa textures found near the stability wall do not necessarily
persist into the interior of the stable region. However, as we now
show, the Yukawa textures {\it do persist}. To prove this, we use the
notion of holomorphy.

\subsection{Holomorphy of the Superpotential and General Textures}\label{ancont}

For a generic heterotic compactification which preserves ${\cal{N}}=1$
supersymmetry, has an $E_6$ GUT factor in its four dimensional gauge
group, and has vanishing cosmological constant, any matter
superpotential Yukawa coupling in the effective low-energy theory is
of the form $\lambda F^{3}$, where $F$ is either a ${\bf {27}}$ or a
${\overline{{\bf{ 27}}}}$ of $E_{6}$. Furthermore, each coefficient
\begin{equation}
\lambda=\lambda ( \left<C_{i} \right>, \left< \mathfrak{z}^a\right>, \left< \phi \right>)
\label{try1}
\end{equation}
must be a holomorphic function on the complex vacuum manifold
${\cal{M}}$ of flat directions of the effective potential energy. Note
that these couplings only depend upon the $C$ and $\phi$ fields, which
we have already encountered, and the complex structure moduli of the
Calabi-Yau threefold, $\mathfrak{z}^a$. Now consider the following general
theorem.
\begin{itemize}
\item {\it If a multivariate holomorphic function with domain $U \subset \mathbb{C}^n$
vanishes on an open subset $B \subset U$, then it vanishes everywhere on
$U$} \cite{jltaylor}. 
\end{itemize}
Let us identify $U$ with a patch in an open cover of ${\cal{M}}$ such
that it contains $B$, an open subset, which covers a region on and
near to the stability wall.  We know from the preceding discussion
that the coupling parameters of $F_{1}^{3}$ and ${\bar{F}}_{2}^{3}$
are both holomorphic functions on ${\cal{M}}$ which do indeed vanish
on such an open patch near the wall. By covering the vacuum space
${\cal{M}}$ with open patches, overlapping on open intersections, we
see from the above theorem that both the $F_1^3$ and
${\bar{F}}_{2}^{3}$ couplings must vanish everywhere - that is, they
vanish identically in the complete vacuum space, not just near the
stability wall. An open cover of this form can be found on any smooth
manifold. On the other hand, any Yukawa couplings, such as $F_{1}
F_{2}^{2}$ or $F_{2}^{3}$, whose holomorphic parameters do {\it not}
vanish in an open region near the wall, will not vanish anywhere in
the interior of the stable chamber with the possible exception of
isolated regions of higher co-dimension. We conclude that: {\it Yukawa
  textures appearing near the stability wall due to invariance under
  the extended $U(1)$ charge, persist throughout the entire stable
  region, arbitrarily far from the wall, even though the $U(1)$ has
  been integrated out of the theory. This result follows simply from
  the holomorphicity of the superpotential.}

We have been considering the branch of the vacuum where, near the wall, $\left<C_2 \right> \neq 0$ and $\left< C_1
\right>=0$. In general, as discussed in Subsection \ref{csection}, there is second branch
defined by  \eqref{branch_bundles2}, where $\mu({\cal
  K})<0$ (i.e. $\mu(\cF)>0$), $\left< C_1 \right> \neq0$ and $\left< C_2
\right>=0$
In this second branch, we see from
\eqref{Wgeneral} that there can be non-vanishing Yukawa couplings, such as
$ \lambda_1 \left< C_1^2 \right> F_1^3$ for example, that are absent in
\eqref{Wnearwall}.
How is this compatible with the above claim that if the Yukawa
couplings vanish in an open subset of the vacuum space then they
vanish everywhere? The answer is simply that, while each branch of the
supersymmetric Minkowski vacuum space is a smooth manifold, the locus
where they intersect is not. Such an intersection can not be
covered with open sets 
with open
intersections. In particular, denoting the two branches above by ${\cal{M}}_1$ and
${\cal{M}}_2$, if we take an open set $B_1 \subset {\cal{M}}_1$ and another $B_2 \subset {\cal{M}}_2$
then, if they intersect at all, their intersection must obey the condition $\left <C_1\right>=\left<C_2\right>=0$. That is, their intersection is necessarily
closed. One can, therefore, have a holomorphic function, such as the $F_1^3$ Yukawa
coupling, that is vanishing everywhere on
one branch of the vacuum space and non-zero on the other - there being
no overlapping open sets to ``communicate'' between
the two.
It follows that we have to make our previous conclusion more specific. That is:
{\it In a given branch of the theory, Yukawa textures near a stability
  wall persist in the entire stable chamber {\bf of that branch}. A
  stable region associated with a different vector bundle separated by
  a stability wall, that is, in a different branch of the theory, {\bf
    need not have identical Yukawa textures}}.

\subsection{A Higher-Dimensional Perspective}

Before proceeding, let us analyze from a higher-dimensional
perspective what is happening when Yukawa couplings ``grow back''. 
Begin  on the stability wall. Expand the dimensional reduction ansatz
\eqref{reduction} to include, for example,
the $F$-fields in
Table \ref{table2}. Then 
\bea \label{reduction2}{\cal{A}}_b = A_b + C_1^{p }
\omega_{p\;b}^{(1) x} T^{(1)}_{x} + C_2^{Q} \omega_b^{(2) y}
T^{(2)}_{y} + F^i_1 \omega^{(3)}_{ib} T^{(3)} + F^{\eta}_2
\omega^{(4)z}_{\eta b} T^{(4)}_z + \ldots 
\eea 
The one-forms $\omega$ are all harmonic with respect to 
the connection, built out of the background gauge field $A$,
appropriate to the representation of the gauge group within which it
is valued. Dimensional reduction then determines the Yukawa coupling
parameters as integrals of the cubic product of these forms over the
Calabi-Yau threefold.  For example, the $F_2^3$ Yukawa coupling is proportional too
\bea \label{mrint} \int_X f_{xyz}\; \omega^{(4)x} \wedge \omega^{(4)y}
\wedge \omega^{(4)z} \wedge \overline{\Omega} \ , \eea where $\Omega$
is the holomorphic three-form and $f_{xyz}$ projects the wedge product
of three one forms onto the gauge singlet. Thus, the texture
\eqref{Wwall} we observed on the stability wall arising from the
extended $U(1)$ gauge invariance can be viewed simply as the vanishing
or non-vanishing of such integrals. For example, on the wall integral
\eqref{mrint} vanishes, simply as a consequence of the contraction of
the one forms with $f$.

As one moves away from the stability wall, the $C$ fields must acquire non-zero vevs to cancel the FI term. {\it Near the stability wall}, where these vevs are small, their contribution to the connection
can be dealt with perturbatively. 
Expanding $C_{i}=\left<C_{i}\right>+\delta C_{i}$, instead
of \eqref{reduction2} one could write \bea \label{reduction3}{\cal{A}}_b =
\hat{A}_b + \delta C_1^{p } \omega_{p\;b}^{(1) x} T^{(1)}_{x} + \delta
C_2^{Q} \omega_b^{(2) y} T^{(2)}_{y} + F^i_1 \hat{\omega}^{(3)}_{ib}
T^{(3)} + F^{\eta}_2 \hat{\omega}^{(4)z}_{\eta b} T^{(4)}_z + \ldots \
, \eea where
\begin{equation}
\hat{A}_b=A_b + \left<C_1^{p }\right>
\omega_{p\;b}^{(1) x} T^{(1)}_{x} + \left<C_2^{Q}\right> \omega_b^{(2) y}
T^{(2)}_{y} \ .
\label{ny1}
\end{equation}
The one-forms $\hat{\omega}$ can now be taken to be harmonic with
respect to connections built out of $\hat{A}$ and the Yukawa coupling
parameters become integrals of cubic products of {\it these} forms.
For example, the $F_2^3$ Yukawa coupling is now
\bea \label{mrint2} \int_X f_{xyz} \;
\hat{\omega}^{(4)x} \wedge \hat{\omega}^{(4)y} \wedge \hat{\omega}^{(4)z} \wedge
\overline{\Omega}  . \eea 
Thus, the texture \eqref{Wnearwall} near the stability wall arising from the {\it spontaneous breaking} of extended $U(1)$ gauge invariance can be viewed as the vanishing or non-vanishing of these integrals. As an example,  \eref{mrint2} {\it no longer vanishes}.

What happens deep in the interior of a stable chamber? {\it Far from the stability wall},
the vevs of
the $C$ fields become so large that their contribution to the connection
is comparable to
$A$. At this stage,  perturbative expansion \eqref{ny1} breaks down
and \eqref{reduction2} becomes
\begin{equation}
{\cal{A}}_b = \tilde{A}_b + F^i_1 \tilde{\omega}^{(3)}_{ib} T^{(3)} + F^{\eta}_2
\tilde{\omega}^{(4)}_{\eta b} T^{(4)} + \ldots \ ,
\label{ny2}
\end{equation}
where the one-forms $\tilde{\omega}$ are harmonic with respect to
connections built out of $\tilde{A}$. Unfortunately, the
indecomposable connection $\tilde{A}$ is no longer related to the
reducible connection $A$ on the stability wall via a perturbative
expansion. Hence, {\it a priori} one has no idea what the texture of
the cubic $\tilde{\omega}$ integrals are. Unlike the case near the wall,
texture here cannot be found by inserting the non-zero $C$ vevs into
\eqref{Wgeneral}. However, the analysis of the preceding subsection
shows that a connection between the two theories can indeed be
determined using the holomorphicity of the superpotential.

With these observations on holomorphy and general textures in hand, we turn next to a more complicated, and restrictive, example of wall-induced Yukawa textures.
 \section{One Wall with Two D-Terms}\label{onewall2D} 

 In the previous section, we considered the case of a {\it single} stability
 wall in the K\"ahler cone where the bundle splits into  {\it two}
 pieces. There are two immediate generalizations of this. Here, we
 describe what happens in cases where the bundle splits into {\it more
   than two} pieces on a {\it single} wall. In the next section, we
 discuss the situation where multiple stability walls are present
 inside the K\"ahler cone.  The simplest case where a bundle can split
 into more than two pieces occurs for an $SU(3)$ structure
 group. Therefore, as previously, we will illustrate the main ideas
 using rank three bundles whose K\"ahler cone contains a single
 stability wall. Now, however, the structure group $SU(3)$ decomposes
 into $S[U(1)\times U(1) \times U(1)]$ on this wall. The conclusions
 drawn from this analysis remain unchanged for bundles of any rank.

 Consider an $SU(3)$ bundle $V$ that splits on the stability wall, not
 as $V= {\cal F} \oplus {\cal K}$ as in the previous section, but
 rather as \beq \label{3piece} V= l_1 \oplus l_2 \oplus l_3 \eeq where
 $l_1$,$l_2$, and $l_3$ are line bundles. In the ${\cal F} \oplus {\cal
   K}$ case, ${\cal{F}}$ was the destabilizing sub-bundle, the
 vanishing of whose slope defined the stability wall. For the
 decomposition \eqref{3piece}, let us take a case where two of the
 three line bundles  destabilize $V$. Choosing such a pair of
 line bundles, the stability wall is defined by the simultaneous
 vanishing of their slopes.  On this locus, the structure group
 changes from $SU(3)$ to $S[U(1)\times U(1) \times U(1)] \cong U(1)
 \times U(1)$, where the last relation holds at least locally.
Therefore, the low energy gauge group changes from $E_6$ in the
interior of any stable region to $E_6 \times U(1)
\times U(1)$ on the stability wall, and
we get {\it two} extended Abelian gauge group
factors rather than one.
The analysis of Yukawa textures is similar to Section \ref{yukawawall}. Now, however, there are {\it two} D-terms of the form \eqref{dterm2}, one for each $U(1)$ factor,
and the theory near the stability wall is restricted by {\it both}
extended Abelian symmetries. Despite this, the analytic continuation arguments of Section
\ref{ancont} remain unchanged. Hence, within a given branch, Yukawa
textures near the stability wall persist over that 
entire stable chamber of the K\"ahler cone, arbitrarily far from the wall.

Given \eqref{3piece}, the group theory which determines 
which multiplets can
appear in the four-dimensional theory near the
stability wall is \bea \label{2Ddecomp} E_8 &\supset& E_6 \times U(1) \times U(1)
\\ \nonumber {\bf 248} &=& {\bf 1}_{0,0} + {\bf 1}_{\frac{1}{2},3} +
{\bf 1}_{-\frac{1}{2},3} + {\bf 1}_{\frac{1}{2},-3} + {\bf
  1}_{-\frac{1}{2},-3} + {\bf 1}_{1,0}+ {\bf 1}_{-1,0} + {\bf
  78}_{0,0} \\ \nonumber && + {\bf 27}_{0,-2} + {\bf
  27}_{\frac{1}{2},1} + {\bf 27}_{-\frac{1}{2},1} + \overline{{\bf
    27}}_{0,2} + \overline{{\bf 27}}_{\frac{1}{2},-1} + \overline{{\bf
    27}}_{-\frac{1}{2},-1} \ ,\eea where the bold face number is the
dimension of the $E_6$ representation and the subscripts are the
two $U(1)$ charges.
The multiplicity of each such multiplet is determined by the number of zero-modes 
of the associated six-dimensional Dirac operator. These are
given by the dimensions of  bundle-valued cohomology groups. The representations, field names and the associated cohomologies for a {\it generic} bundle of type \eqref{3piece} are listed in the first three columns of Table \ref{tabledoubled}. Note that there are now {\it six} different types of charged bundle moduli, that is, $C$-fields, of the kind described in Section \ref{csection}. As discussed in that subsection, the $C$ fields are intimately related to the branch structure of the theory. We now generalize the analysis of Section \ref{csection} to the present case.

First choose which two line bundles in \eqref{3piece} destabilize $V$. Let us take $l_{1}$ and $l_{2}$ for specificity. Associated with each will be a D-term of the generic form
\eqref{dterm2}, where $D_{1}^{U(1)}$ and $D_{2}^{U(1)}$ contain the slope $\mu (l_{1})$ 
and $\mu (l_{2})$ respectively. Although both slopes vanish on the stability wall, the assumption that the associated line bundles destabilize $V$ implies that their slopes become negative in the interior of the stable chamber. Now note that, in addition to these two D-terms, one must consider the superpotential. Ignoring the $E_{6}$ non-singlets, this can be written as
\begin{equation}
W= C_1 \tilde{C}_2 \tilde{C_3} + (C_1 \tilde{C}_1)^2+(C_2 \tilde{C}_2)^2 +(C_3
\tilde{C}_3)^2 + C_1 C_2 \tilde{C}_1 \tilde{C}_2 + C_1 C_3 \tilde{C}_1
\tilde{C}_3 + C_2 C_3 \tilde{C}_2 \tilde{C}_3 \;.
\label{xmas1}
\end{equation}
For simplicity, here and in the remainder of the paper, we suppress
indices and the coefficients in front of each term. In addition, we
work only to the dimension required for our analysis. In any stable
region, the four-dimensional effective theory has a supersymmetric
vacuum with vanishing cosmological constant. Therefore, as we vary the
K\"ahler moduli away from the stability wall into the $\mu (l_{1})
<0$, $\mu (l_{2})<0$ region, in addition to the vanishing of the two
D-terms, we must set
\begin{equation}
\partial_{C_{i}}W=\partial_{{\tilde{C}}_{j}}W= W=0 \qquad  i,j=1,2,3 \ .
\label{xmas2}
\end{equation}
Terms of the form $(C_i \tilde{C}_i)^2$, $i=1,2,3$ in
\eqref{xmas1} ensure that,  for each index $i$, either $C_i$ or $\tilde{C}_i$, but not both, can have a non-zero vev. The terms of the form $C_1 \tilde{C}_2 \tilde{C}_3$
and $\tilde{C}_1 C_2 C_3$ in \eqref{xmas1} ensure that only one of
$C_1$, $\tilde{C}_2$ and $\tilde{C}_3$, and only one of $\tilde{C}_1$,
$C_2$ and $C_3$, obtains a non-zero vev.  
Combining these results with the requirement that $D_{1}^{U(1)}=D_{2}^{U(1)}=0$, we find that there are $\it six$ supersymmetric branches associated with this stability wall in the K\"ahler cone - each branch specified by a pair of non-vanishing $C$ fields. For example, one branch is given by
\begin{equation}
\left<\tilde{C}_{2}\right>  \neq 0 \ , \quad  \left<C_{3}\right> \neq 0 
\label{xmas3}
\end{equation}
where all other $\left<C\right>=0$.

In terms of sequences, the different possible $C$ field vevs
correspond to the different ways of building a bundle $V$
from the three constituent line bundles $l_1$, $l_2$ and $l_3$. Let us
take the case \eqref{xmas3}, where $\tilde{C}_2$ and $C_3$ have non-zero vevs, as a specific example.
Consider the two sequences
\bea
0 \to l_1 \to {\cal{W}} \to l_3 \to 0 \label{firstone} \\
0 \to l_2 \to V \to {\cal{W}} \to 0 \label{secondone}
\eea
The moduli space of the first sequence is described by $Ext^1(l_3,l_1)
\cong H^1 (l_1 \otimes l_3^*)$. Therefore, this extension is non-trivial,
that is,  ${\cal{W}} \neq l_1 \oplus l_3$, if and only if  one is at a non-zero element of
this cohomology group. We see from Table \ref{tabledoubled} that this
corresponds, in field theory language, to $< \tilde{C}_2> \neq 0$ .
Sequence \eqref{secondone} is a non-trivial extension if and only if one is 
at a non-trivial element in $Ext^1({\cal{W}},l_2) \cong H^1(l_2 \otimes
{\cal{W}}^*)$. Using the dual sequence to \eqref{firstone}, we find \bea \label{tensored} 0 \to
l_2 \otimes l_3^* \to l_2 \otimes {\cal{W}}^* \to l_2 \otimes l_1^* \to 0 \;.
\eea Since all vevs for the $\tilde{C}_1$ fields vanish,
it follows from Table \ref{tabledoubled} that this branch is confined to the zero-element
of $H^1(l_2 \otimes l_1^*)$. The long exact
sequence associated with \eqref{tensored} then simplifies to  
\bea \ldots \to H^1(l_2
\otimes l_3^*) \to H^1(l_2 \otimes {\cal{W}}^*) \to 0 \ . \eea 
It follows that any non-zero element of $H^1(l_2 \otimes
l_3^*)$, the cohomology associated with the fields $C_3$, maps to a non-zero element of $H^1(l_2 \otimes {\cal{W}}^*)$, the
cohomology associated with bundle \eqref{secondone}. 
That is, the deviation of the bundle, $V$, away from its split point in
sequence \eqref{secondone} is controlled by the $\left< C_3 \right>
\neq 0$ condition in the field theory. 
Putting everything together, we conclude that $V$ in
\eqref{secondone} is indeed the bundle corresponding to the branch
of the vacuum space where
$\left< \tilde{C}_2 \right> \neq 0,\left < C_3 \right> \neq 0$ and all other $C$ vevs vanish. A  similar analysis can be
performed for any other allowed branch.

We now turn to an analysis of the allowed Yukawa textures. All of the fields in Table \ref{tabledoubled}, if present in a specific example, are
massless near the wall. 
\begin{table}[h]
\begin{center}
\begin{tabular}{|c|c|c||c|}
  \hline
  Representation & Field Name & Cohomology & Multiplicity \\ \hline
  ${\bf 1}_{\frac{1}{2},3}$ & $C_1$  & $h^1(X,l_1^* \otimes l_2)$& 0\\ \hline 
  ${\bf 1}_{-\frac{1}{2},-3}$ & $\tilde{C}_1$ &  $h^1(X,l_1 \otimes l_2^*)$& 0\\ \hline  
  ${\bf 1}_{-\frac{1}{2},3}$ & $C_2$ &  $h^1(X, l_1^* \otimes l_3)$  & 0\\ \hline
  ${\bf 1}_{\frac{1}{2},-3}$ & $\tilde{C}_2$ & $h^1(X,l_1 \otimes l_3^*)$ & 100\\ \hline 
  ${\bf 1}_{1,0}$ & $C_3$  & $h^1(X,l_2 \otimes l_3^*)$ & 200\\ \hline
  ${\bf 1}_{-1,0}$ & $\tilde{C}_3$  & $h^1(X,l_2^* \otimes l_3)$ & 0\\ \hline  
  ${\bf 27}_{0,-2}$ & $f_1$  & $h^1(X,l_1)$ & 0\\ \hline 
  ${\bf 27}_{\frac{1}{2},1}$ & $f_2$ & $h^1(X,l_2)$ & 20\\ \hline 
  ${\bf 27}_{-\frac{1}{2},1}$ & $f_3$  &  $h^1(X,l_3)$& 0\\ \hline
  $\overline{{\bf 27}}_{0,2}$ &  $\tilde{f}_1$  &  $h^1(X,l_1^*)$& 0 \\ \hline 
  $\overline{{\bf 27}}_{-\frac{1}{2},-1}$ & $\tilde{f}_2$ &  $h^1(X,l_2^*)$&0 \\ \hline 
  $\overline{{\bf 27}}_{\frac{1}{2},-1}$ & $\tilde{f}_3$  & $h^1(X,l_3^*)$ & 40\\ \hline 
  \end{tabular}
\mycaption{The representations, field content and cohomologies of a {\it generic} $E_6 \times U(1) \times U(1)$ theory associated with a poly-stable bundle $V= l_1 \oplus l_2 \oplus l_3 $ on the stability wall.
The multiplicities for the {\it explicit} bundle defined by \eqref{wow} are given in the fourth column.}
\label{tabledoubled}
\end{center}
\end{table}
The most general superpotential for cubic matter interactions
invariant under the $E_6 \times U(1) \times U(1)$ symmetry, including
the purely $C$ field superpotential in \eqref{xmas1}, is given by
\bea \nonumber W &=& f_1 f_2 f_3+ C_1 \tilde{C}_2 \tilde{C_3} \\
\nonumber && + (C_1 \tilde{C}_1)^2+(C_2 \tilde{C}_2)^2 +(C_3
\tilde{C}_3)^2 + C_1 C_2 \tilde{C}_1 \tilde{C}_2 + C_1 C_3 \tilde{C}_1
\tilde{C}_3 + C_2 C_3 \tilde{C}_2 \tilde{C}_3 \\ \nonumber && + f_1^2
f_2 C_2 + f_1^2 f_3 C_1 + f_2^2 f_1 \tilde{C}_3 + f_2^2 f_3
\tilde{C}_1 + f_3^2 f_1 C_3 + f_3^2 f_2 \tilde{C}_2 \\ \label{2Dw} &&
+f_1^3 C_1 C_2 + f_1^2 f_2 C_1 \tilde{C}_3 + f_1^2 f_3 C_2 C_3 + f_1
f_2^2 \tilde{C}_1 C_2+ f_1 f_3^2 C_1 \tilde{C}_2 \\ \nonumber &&+
f_2^3 \tilde{C}_1 \tilde{C}_3 + f_2^2 f_3 \tilde{C}_2 \tilde{C}_3 +
f_2 f_3^2 \tilde{C}_1 C_3 + f_3^3 \tilde{C}_2 C_3 \\ \nonumber && +
f_1^3 C_1^2 \tilde{C}_3 + f_1^3 C_2^2 C_3 + f_2^3 \tilde{C}_1^2 C_2 +
f_2^3 \tilde{C}_2 \tilde{C}_3^2 + f_3^3 C_1 \tilde{C}_2^2 + f_3^3
\tilde{C}_1 C_3^2 \\ \nonumber && + \; \; \;
\{f_1,f_2,f_3,C_1,C_2,C_3\} \leftrightarrow \{
\tilde{f}_1,\tilde{f}_2,\tilde{f}_3,\tilde{C}_1,\tilde{C}_2,\tilde{C}_3
\} \ , \eea where terms are shown in the order of increasing dimension
and we do not display any coefficients or indices.  No quadratic terms
appear, since all these superfields are zero-modes of the
compactification. Furthermore, interactions of dimension six or higher
in $E_6$ non-singlet fields $f$ and $\tilde{f}$ are not relevant to
the discussion, so we ignore them. Finally, displayed are the lowest
dimension terms required in our analysis. Each interaction in
\eqref{2Dw} can be multiplied by any positive integer power of neutral
combinations of $C$ fields. These do not change the subsequent
analysis and, hence, in the interests of brevity, we suppress them.

Examining \eqref{2Dw}, we see that the only Yukawa couplings present {\it on the stability wall}, where all
$C$ field vevs vanish, are 
\begin{equation}
W_{\rm Yukawa}^{\rm wall}=f_1 f_2 f_3+\tilde{f}_1 \tilde{f}_2 \tilde{f}_3 \ .
\label{try2}
\end{equation}
This is a very restrictive texture. As in the previous section, some of
the missing Yukawa couplings can ``grow back'' as one moves away from the
stability wall into a stable chamber of the K\"ahler
cone. 
Returning to \eqref{2Dw}, it is clear that there are several
possible Yukawa textures that can result from the
the splitting of an $SU(3)$ bundle into three line bundles on the stability wall. Which texture occurs depends on which $C$ fields get non-zero vevs, that is, which branch of the theory one is on. For the representative branch discussed above, where
$\left <\tilde{C}_2\right> \neq 0, \left<C_3\right > \neq 0$ and all other $C$ vevs vanish, we find that
\bea \label{specialcase} W_{\textnormal{Yukawa}}^{\rm near \ wall} &=& f_1 f_2 f_3 + \tilde{f}_1 \tilde{f}_2 \tilde{f}_3+ {\bf  \left< C_3 \right>f_3^2
f_1+  \left< \tilde{C}_2 \right>f_3^2 f_2}\\ \nonumber && {\bf +  \left< \tilde{C}_2 \right> \tilde{f}_1^2 \tilde{f}_2 + \left< C_3 \right> \tilde{f}_2^2 \tilde{f}_1 +\left< \tilde{C}_2\right>  \left< C_3 \right>f_3^3} \;.  \eea 

Comparing to \eqref{try2}, it follows that there are five different types of Yukawa couplings which can ``grow back'' as we deform to the indecomposable bundle described by
\eqref{firstone} and \eqref{secondone}. These are shown in boldface. Be this as it may, it is important to note that there remain {\it many} Yukawa couplings, such as $f_{1}^{3}, \ \tilde{f}_{2}^{3},  \ f_{2}\tilde{f}_{1}^{2}$, which are forbidden by the the extended $U(1) \times U(1)$ symmetry. Finally, using the holomorphy analysis from subsection \ref{ancont}, we conclude that {\it everywhere} in this stable chamber the Yukawa texture is given by 
\bea \label{specialcase2} W_{\textnormal{Yukawa}} = f_1 f_2 f_3 + \tilde{f}_1 \tilde{f}_2 \tilde{f}_3+ {\bf  f_3^2
f_1+  f_3^2 f_2}{\bf +  \tilde{f}_1^2 \tilde{f}_2 + \tilde{f}_2^2 \tilde{f}_1 +f_3^3} \;.  \eea 

\subsection*{An Example}
As an example of a stability wall of the type discussed
in this section, consider the bundle
\bea \label{2dext}
0 \to {\cal O}(-1,1) \oplus {\cal O}(-2,2) \to V \to {\cal O}(3,-3) \to 0
\eea
defined on the complete intersection Calabi-Yau threefold
\begin{eqnarray}
\label{eg2cy3}
\left[ \ba{c |c }
\mathbb{P}^1 & 2 \\
\mathbb{P}^3 & 4
\ea \right]^{2,86} \ .
\end{eqnarray}
Equation \eqref{2dext} describes $V$ as an extension of direct sums of
line bundles. This bundle has a stability wall on the line of slope
one in its two-dimensional K\"ahler cone. On this locus, the bundle
splits as 
\bea V = {\cal O}(-1,1) \oplus {\cal O}(-2,2) \oplus {\cal
  O}(3,-3)\;. \label{wow} \eea 
Hence, we can identify $l_1$, $l_2$ and $l_3$ of the
previous discussion as ${\cal O}(-1,1)$, ${\cal O}(-2,2)$ and ${\cal
  O}(3,-3)$ respectively.

Given this explicit example, one can calculate the multiplicity of each
multiplet described in \eqref{2Ddecomp}. These are presented in the fourth
column of Table \ref{tabledoubled}. Note that there are
20 $f_2$ fields and no other $\bf{27}$ multiplets of $E_6$. Additionally, there are 40
$\tilde{f}_3$ fields but no other $\overline{\bf {27}}$  anti-generations. Importantly, the only $C$ fields which appear are $\tilde{C}_2$ and $C_3$, precisely the fields that got non-zero vevs in our preceding discussion. Combining
this information with \eqref{specialcase2}, we find that there are {\it no Yukawa couplings at
all}, either between three ${\bf 27}$'s or between three
$\overline{{\bf 27}}$'s. The residual symmetries left over in the
interior of this stable region, as a result of the presence of the stability
wall, completely remove all couplings between the matter families in this
example. This is a good illustration of the importance of residual
symmetries. In some cases they can be extremely restrictive, and could forbid some, or all, of the interactions required by phenomenology.
  
  
\section{Two Walls in the K\"ahler Cone}\label{2walls}

\begin{figure}[!ht]
  \centerline{\epsfxsize=5in\epsfbox{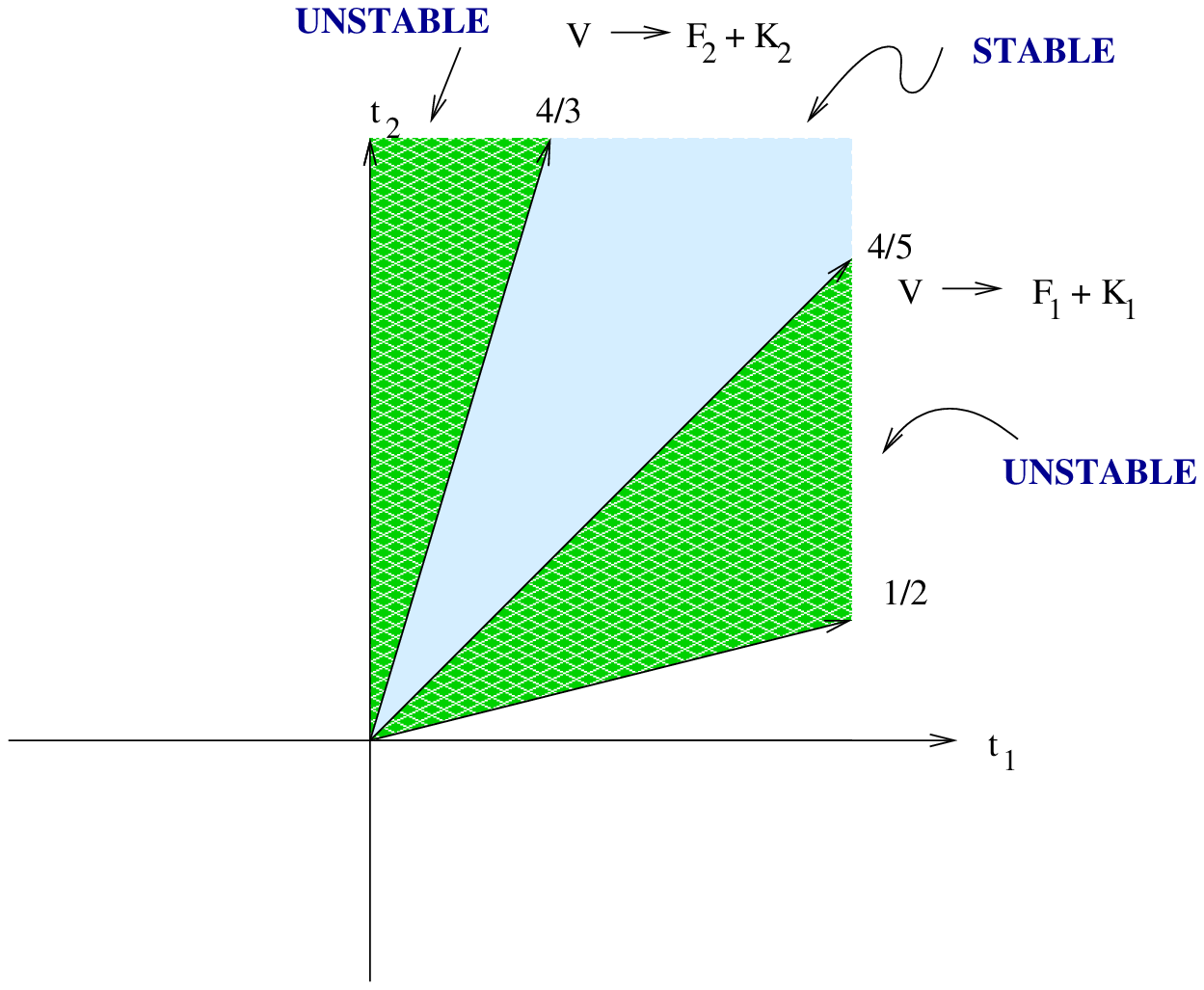}} \mycaption{
    The K\"ahler cone (The set of moduli $t^1,t^2>0$ and $2t^2>t^1$) and the regions of stability/instability for the
    Calabi-Yau threefold \eqref{twofour} and the bundle \eqref{doublewalleg}. 
    At the lower boundary, $V$ decomposes as $V \rightarrow \cF_1 \oplus {\cal K}_1$, where $\cF_1$ is defined in \eref{F1}. At the upper boundary, the poly-stable decomposition is given by $V \rightarrow \cF_2 \oplus {\cal K}_2$, with $\cF_2$ defined by \eref{F2}.}
\label{doublewallfig}
\end{figure}

In Section \ref{yukawawall}, we considered a single stability wall in
the K\"ahler cone where the bundle split into two pieces.  This was
generalized in the preceding section to the case of a bundle which
split into three or more pieces, again on a single stability
wall. More generally, however, a supersymmetric chamber in the
K\"ahler cone can be surrounded by multiple stability walls.  In
this section, we turn our attention to this situation.
The simplest examples occur in vacua with $h^{1,1}=2$. In this case,
there can be at most two stability walls bounding a supersymmetric
region (see Figure \ref{doublewallfig}). In the most basic examples the vector bundle splits into just
two pieces on each wall. We will restrict our discussion to rank three
cases in order to illustrate these multi-wall scenarios, and the
Yukawa textures they give rise to.  We emphasize, however, that the
general type of conclusions drawn from this analysis remain unchanged
for $h^{1,1} \geq 3$, vector bundles of any rank, and for more general
decompositions of the bundle. 

Clearly, the analysis of Sections \ref{section1} and \ref{yukawawall}
applies to each of the two stability walls individually. Each wall,
therefore, places constraints on the terms in the four-dimensional
theory. The question of exactly how these constraints interrelate,
however, is not easily answered. Note that the description of the
effective theory, including the labeling of the fields, and possibly
even the number of vector-like pairs, changes between boundaries.  For
example, consider a vacuum which, in the interior of the stable
region, has three chiral $\bf {27}$ matter families of $E_6$.
Furthermore, assume that at the ``upper" stability wall no
family/anti-family pairs appear and that two families get charge $q_1$
and one family charge $Q_1$ under the extended $U(1)$ symmetry. Now
suppose that at the ``lower" boundary this second stability wall gives
two families of charge $q_2$ and one of charge $Q_2$ under its extra
Abelian gauge group. In general then, every field in the problem can
carry two additional quantum numbers, one associated with the $U(1)$
at the upper boundary and the second with the Abelian symmetry at the
lower boundary. Each $U(1)$ gauge symmetry only appears near the
stability wall which gives rise to it, and so the associated charges
only have meaning in that part of field space. We have two ``near
wall'' theories with no overlapping region of validity. Given this,
when we consider the fields at a generic point in the slope-stable
region, how do we correlate the charges which they acquire as we near
each of the two walls? For example, do we have two charged objects
which pick up charge $q_1$ near one wall and $q_2$ near the other, or
perhaps simply one and some fields with charges $Q_1$ and $q_2$?  As
may be expected, answering this type of question and, hence,
describing the physics of multiple stability walls is, in general,
example dependent. To untangle the most general constraints on the
theory requires a careful observation of the chosen fields and
geometry. There are some cases where the result is particularly
simple, however, and we will give an illustrative example here.

To begin, consider the complete intersection manifold
\bea \label{twofour}
X=\left[ \ba{c|c}
\mathbb{P}^1 & 2 \\
{\mathbb P}^3 & 4 
\ea
\right]^{2,86} \ . \eea
Over this space, we construct the following $SU(3)$ monad bundle,
\beq \label{doublewalleg} 0 \to V \to {\cal O}(1,1) \oplus {\cal
  O}(2,0) \oplus {\cal O}(3,-1) \oplus {\cal O}(-2,1) \to {\cal
  O}(4,1) \to 0 \ . \eeq
A stability analysis as in \cite{Anderson:2009nt,Anderson:2008ex,stability_paper} reveals that this bundle is destabilized by a pair of rank two sub-bundles, namely

\beq \label{F1} 0 \to {\cal F}_1 \to {\cal O}(1,1) \oplus {\cal
  O}(2,0) \oplus {\cal O}(-2,1) \to {\cal O}(4,1) \to 0 \eeq where
$c_1(\cF_1)=(-3,1)$ and \beq \label{F2} 0 \to {\cal F}_2 \to {\cal
  O}(1,1) \oplus {\cal O}(2,0) \oplus {\cal O}(3,-1) \to {\cal O}(4,1)
\to 0 \eeq with $c_1(\cF_2)=(2,-1)$.  The vanishing of the slope of
the first of these sub-bundles, $\cF_1$, provides a ``lower" boundary
wall to the stable region of K\"ahler moduli space, while the
vanishing of the slope of the second, $\cF_2$, provides an ``upper"
boundary to this region. The stability wall structure for this example
is given in Figure \ref{doublewallfig}. We will consider each boundary
in turn, and the effective field theory associated with it, before
combining our observations to find the constraints on the full theory
at a generic point in the stable region.

\begin{table}[t]
\begin{center}
\begin{tabular}{|c|c|c|c|}
  \hline
  Representation & Field Name & Cohomology & Multiplicity \\ \hline
  $({\bf 27},{\bf 2})_{-1}$ & $f_1$ & $H^1(X,{\cal F}_1)$ & 13 \\ \hline
  $({\bf \overline{27}},{\bf 2})_{1}$ & $f_3$ & $H^1(X,{\cal F}_{1}^{*})$ & 1 \\ \hline
  $({\bf \overline{27}},{\bf 1})_{-2}$ & $f_4$ & $H^1(X,{\cal K}_{1}^{*})$ & 8 \\ \hline
  $({\bf 1},{\bf 2})_{-3}$ & $C_1$ & $H^1(X,{\cal F}_1 \otimes {\cal K}_{1}^{*})$ & 70 \\ \hline
  \end{tabular}
\mycaption{Field content on the lower
  stability wall of the {\it explicit} bundle \eqref{doublewalleg}. Cohomologies with vanishing multiplicity are not shown.}
\label{table3}
\end{center}
\end{table}

\begin{table}[h]
\begin{center}
\begin{tabular}{|c|c|c|c|}
  \hline
  Representation & Field Name & Cohomology & Multiplicity \\ \hline
  $({\bf 27},{\bf 2})_{-1}$ & $\tilde{f}_1$ & $H^1(X,{\cal F}_2)$ & 9 \\ \hline
  $({\bf \overline{27}},{\bf 2})_{1}$ & $\tilde{f}_3$ & $H^1(X,{\cal F}_{2}^{*})$ & 9 \\ \hline
  $({\bf 27},{\bf 1})_{2}$ & $\tilde{f}_2$ & $H^1(X,{\cal K}_{2})$ & 4 \\ \hline
  $({\bf 1},{\bf 2})_{-3}$ & $\tilde{C}_1$ & $H^1(X,{\cal F}_2 \otimes {\cal K}_{2}^{*})$ & 3 \\ \hline
  $({\bf 1},{\bf 2})_{3}$ & $\tilde{C}_2$ & $H^1(X,{\cal F}_{2}^{*} \otimes {\cal K}_{2})$ & 49 \\ \hline
  \end{tabular}
\mycaption{Field content on the upper
  stability wall of {\it explicit} bundle \eqref{doublewalleg}. Cohomologies with vanishing multiplicity are not shown.}
\label{table4}
\end{center}
\end{table}

Let us begin our analysis on the ``lower' boundary wall defined by the sub-bundle $\cF_1$. For the theory to be supersymmetric, the bundle $V$ in \eref{doublewalleg} must ``split" on this stability wall into the direct sum $\cF_1 \oplus {\cal K}_1$, where ${\cal K}_1 =\cO(3,-1)$. The field content near this 
boundary, and the charges under the extended $E_{6} \times U(1)$ gauge symmetry, are given in Table \ref{table3}. To third order in the matter fields, the invariant superpotential is 
\beq\label{w_high}
W= f_3^{2}f_4+f_3^{3}C_1 \ .
\eeq
As always, we ignore irrelevant higher dimension terms.
Note that on the lower stability wall, all ${\bf 27}^3$ Yukawa couplings are forbidden entirely. Furthermore,  the ${\bf \overline{27}}^3$ couplings exhibit a very restrictive texture. For example, the $f_{i}^{3}$, $i=1,2,3$ terms are absent. What happens for small deformations away from this wall into the stable chamber?
Since one can describe the stable bundle $V$ in terms of this de-stabilizing sub-bundle as
\beq\label{extf1}
0 \to \cF_1 \to V \to {\cal K}_1 \to 0 \ ,
\eeq
we see that $C_1 \in H^1 (X, \cF_1 \times {\cal K}^{*}_1)$ must acquire a non-zero vev in order to cancel the FI piece of the D-term
associated with the upper boundary, see \eref{dterm}. As a result, the ${\bf \overline{27}}^3$ Yukawa coupling $\left<C_1\right>f_3^{3}$ can ``grow back'' near this stability wall. It follows from the holomorphy analysis of subsection \ref{ancont} that one expects
\begin{equation}
W_{\rm Yukawa}= f_3^{2}f_4+{\bf f_3^{3}}
 \label{home1}
 \end{equation}
{\it everywhere} in the interior of the stable chamber.

Turn now to the upper boundary, where the theory is defined by sub-bundle $\cF_2$ in \eref{F2}. The polystable decomposition is now $V\rightarrow \cF_2 \oplus {\cal K}_2$, with ${\cal K}_{2}=\cO(-2,1)$. This is associated again with an extended $E_{6} \times U(1)$ symmetry of the four-dimensional theory. However, the $U(1)$ symmetry near this wall is {\it not} the same as the Abelian symmetry on the lower boundary. Near the upper boundary, the field content and charges under the {\it new} extended $E_{6} \times U(1)$ gauge symmetry are listed in Table \ref{table4}.
The relevant superpotential is now given by
\beq\label{W_low}
W = \tilde{f}_{1}^{2}\tilde{f}_2+\tilde{f}_{2}^{2}\tilde{f}_{1}\tilde{C}_1+\tilde{f}_1^3\tilde{C}_2+\tilde{f}_3^3\tilde{C}_1+\tilde{f}_{2}^3\tilde{C}_{1}^{2}+\tilde{f}_{1}^{2}\tilde{f}_{2}\tilde{C}_1\tilde{C}_{2}+(\tilde{C}_{1}\tilde{C}_2)^2 \ .
\eeq
Note that {\it on} the upper stability wall, the ${\bf{27}}^{3}$ Yukawa couplings exhibit a very restrictive texture. Furthermore, all $\overline{\bf{27}}^{3}$ terms are disallowed. What happens for small deformations away from this wall into the stable chamber?

Note that while both charged moduli $\tilde{C}_1,\tilde{C}_2$ are
present (recall that these are the moduli responsible
for ``re-mixing" $\cF_2 \oplus {\cal K}_2$, into
$V$ in \eref{doublewalleg}), as discussed in Section \ref{csection},
only one of them can get a vev in the stable region. Since $\cF_2$ in
$V$ is the destabilizing sub-bundle at the upper boundary, it is clear
that we can describe $V$ in the stable region as \beq\label{extF2} 0
\to \cF_2 \to V \to {\cal K}_2 \to 0 \ .  \eeq As a result, it is
$\tilde{C}_1 \in H^1(X, \cF_2 \times {\cal K}_{2}^{*})$ which controls
the movement away from the upper stability wall into the
indecomposable gauge configuration. This can also be seen by
inspecting the charges of the various fields in the $U(1)$ D-term
associated with the upper boundary, see \eref{dterm2}. It follows from
the slope of $\cF_2$ in Figure \ref{doublewallfig} that it is the
$\tilde{C}_1$ fields that must acquire a non-zero vev.  Furthermore,
the $(\tilde{C}_{1}\tilde{C}_2)^2$ term in \eqref{W_low} assures that
the vevs of $\tilde{C}_2$ must be zero in the stable region. As a
result, of the five $C$ field dependent matter couplings in
\eqref{W_low}, three ``grow back'' to contribute to Yukawa couplings
near this stability wall.
 It follows from the holomorphy analysis of subsection \ref{ancont} that one expects
 \begin{equation}
W_{\rm Yukawa} = \tilde{f}_{1}^{2}\tilde{f}_2+{\bf \tilde{f}_{2}^{2}\tilde{f}_{1}+\tilde{f}_3^3+\tilde{f}_{2}^3}
 \label{home2}
 \end{equation}
{\it everywhere} in the interior of the stable chamber.

Now consider the theory deep in the stable region, away from these two boundaries. 
Using \eref{doublewalleg}, one can find the spectrum of the ``standard'' heterotic compactification in the stable region. Since $V$ is a stable $SU(3)$ bundle, $H^0(X, V)=H^3(X,V)=0$ \cite{Green:1987mn,Distler:1987ee}. It follows that the long exact sequence in cohomology  associated with \eref{doublewalleg} splits into
\beq
0 \to H^0(X,\cO(1,1)\oplus \cO(2,0)) \to H^0(X,\cO(4,1)) \to H^1(X,V) \to H^1(X, \cO(-2,1)) \to 0 
\label{home3}
\end{equation}
and
\begin{equation}
0 \to H^2(X,V) \to H^2(X,\cO(3,-1)\oplus \cO(2,0)) \to 0 \ .
\label{home4}
\eeq
As a result, we see that
\beq
h^1(X,V)=h^0(X,\cO(4,1)) -h^0(X,\cO(1,1)\oplus \cO(2,0))+h^1(X, \cO(-2,1))=20-11+4=13 
\label{home5}
\end{equation}
and
\begin{equation}
h^1(X,V^*)=h^2(X,V)= h^2(X,\cO(3,-1)\oplus \cO(2,0))=9
\label{home6}
\eeq
respectively. Hence, in the interior of the stable chamber there are thirteen ${\bf{27}}$ and nine $\overline{{\bf{27}}}$ multiplets of $E_{6}$;  that is, four chiral and nine vector-like pairs of matter families. 
How then does this general theory relate to the effective theories at each boundary wall? To answer this, consider the alternative descriptions of $V$ in terms of $\cF_1$, \eqref{extf1}, and in terms of $\cF_2$, \eqref{extF2}. From the associated long exact cohomology sequences, we find that in the stable region
\begin{eqnarray}
&& \ H^1(X,V)=H^1(X,{\cal F}_1)=H^1(X,{\cal F}_2) \oplus H^1(X,{\cal K}_{2}),  \label{prehome7}\\
&& H^1(X,V^*)=H^1(X,{\cal F}_{1}^{*}) \oplus H^1(X,{\cal K}_{1}^{*})=H^1(X,{\cal F}_{2}^{*}) \ .
\label{home7}
\end{eqnarray}
It then follows from Tables \ref{table3} and \ref{table4} that 
\bea
&& \ H^1(X,V)={\rm span}\{f_1\}={\rm span}\{\tilde{f}_1,\tilde{f}_2\}~~~~~13=(9+4) \ , \label{matching}\\
&& H^1(X,V^*)={\rm span}\{f_3,f_4\}={\rm span}\{\tilde{f}_3\}~~~~~(1+8)=9 \ . \label{matching2}
\eea
The key point in this example is that, on the lower stability wall all of the families acquire the same charge. Equally, on the upper stability wall all of the anti-families acquire the same charge. Thus, we are able to correlate the charges picked up by the  matter fields at the two different walls in an unambiguous manner! Note that the number of chiral families and the number of vector-like pairs stays the same throughout moduli space. Using this result, one can now impose the constraints from {\it each} stability wall on couplings throughout the {\it entire} K\"ahler cone.

First observe from \eqref{home1} and \eqref{matching} that the constraints from the lower wall completely {\it forbid} any ${\bf 27}^3$ Yukawa couplings in the stable chamber. It then follows from \eqref{matching} that the couplings $\tilde{f}_{1}^{2}\tilde{f}_2$, ${\bf {\tilde{f}}_{2}^{2}\tilde{f}_{1}}$ and ${\bf{\tilde{f}}_{2}^{3}}$ in \eref{home2}, while not forbidden by gauge invariance at the upper boundary, are none-the-less vanishing everywhere due to the constraints from the lower wall.
Second, observe from \eqref{home1} and \eqref{matching2} that the lower wall constraints do allow $\overline{{\bf 27}}^3$ Yukawa couplings in the stable chamber, but only in a {\it specific texture with nine terms}. We see from \eqref{matching2} that this is group theoretically consistent with the existence of the ${\bf{\tilde{f}}_{3}^{3}}$ in \eref{home2}. However, the gauge symmetry of the upper wall would allow $\binom{9}{3}+9(9-1)+9=165$ such terms. It follows that additional texture is imposed by the constraints of the lower wall to reduce this number to 9.
Note from \eqref{home1} and \eqref{home2} that these nine holomorphic parameters, while non-vanishing in the interior of the stable region, must depend on bundle moduli in such a way that they all go to zero at the upper boundary, while eight remain non-zero at the lower wall.

We conclude that it is possible to trace the constraints from both boundary stability walls into the interior of the stable chamber. At a generic point of this four-generation, nine vector-like pair $E_6$ theory, we find that there are no ${\bf 27}^3$ couplings allowed and only $9$ specific ${\bf \overline{27}}^3$ Yukawa couplings surviving out of $165$. Note that the stronger constraints arise from the bottom stability wall. Had one only considered constraints from the upper stability wall, a decidedly incomplete texture of Yukawa interactions would have been obtained. This is a clear example of why one must be careful to take into account {\it all} stability walls of the bundle when attempting to determine the Yukawa texture of a heterotic vacuum. 

\section{Textures in a Three Generation Model}\label{threegen}

\subsection{One Heavy Family and Other Textures}\label{phenotext}

Previously, we investigated Yukawa textures arising from stability walls of $SU(3)$ bundles. In this section, we discuss stability walls and their constraint on matter textures in a {\it more phenomenologically realistic} context. Specifically, we consider the $SO(10)$ theory associated with an $SU(4)$ bundle. We further assume that this bundle is destabilized by a single rank two sub-bundle which gives rise to a {\it single} stability wall in the K\"ahler cone and a {\it single} D-term.

In the stable chamber, the structure group of $V$ is $SU(4)$ and, hence, the low energy theory has a gauged $SO(10)$ symmetry. As in previous sections, along the wall of poly-stability the vector bundle splits into a direct sum
\beq\label{so10split}
V \rightarrow \cF_1 \oplus \cF_2 \ ,
\eeq
where now both sub-bundles have rank two.
The structure group then changes from $SU(4)$ to $S[U(2)\times U(2)]$. Since the commutant of $S[U(2)\times U(2)]$ in $E_8$ is $SO(10)\times U(1)$, the symmetry of the four-dimensional theory is enhanced by an anomalous $U(1)$. On the stability wall, where $V$ decomposes as \eref{so10split}, the fields carry an extra $U(1)$ charge in addition to their $SO(10)$ content. We present  the generic zero-mode spectrum in Table \ref{tableso10sec}.

Our goal is to illustrate how the stability wall can constrain Yukawa textures in a phenomenologically realistic  context. Therefore, we will only consider bundles leading to {\it three generations} of chiral matter. To simplify the analysis, these bundles will be further restricted so that the multiplicities of the ${\bf 16}$ and ${\bf 10}$ fields on the stability wall, and, hence, in any of its branches, are
\beq\label{samplematter}
\ba
{l}
n_{16_{+1}}=h^1(X,\cF_1)=2 \ , \quad n_{16_{-1}}=h^1(X,\cF_2)=1 \\
\qquad \qquad \qquad n_{10_{+2}}=h^1(X,\wedge^2 \cF_2)=1
\ea
\eeq
and no other $SO(10)$ non-singlets occur. 
The theory generically contains both $C_1 \in H^1(X,\cF_1 \times \cF^{*}_2)$ with charge $+2$ as well as $C_2 \in H^1(X,\cF_2 \times \cF_{1}^{*})$ with charge $-2$. 
To cubic order in the matter fields, the $SO(10)\times U(1)$ invariant superpotential is
\beq\label{heirarchyeg}
W = h_1f_{2}^{2} +h_1f_{1}f_2C_2+h_{1}f_{1}^{2}C_{2}^{2} +(C_{1}C_{2})^{2} \ .
\eeq
As always, we ignore irrelevant higher dimension terms. {\it On the stability wall},
the FI piece of the associated D-term vanishes. To have an ${\cal{N}}=1$ supersymmetric Minkowski vacuum, if follows from \eqref{dterm2} and \eqref{heirarchyeg} that $\left<C_{1}\right>=\left<C_{2}\right>=0$. Therefore,
\begin{equation}
W^{\rm wall}_{\rm Yukawa} = h_1f_{2}^{2} \ .
\label{sun1}
\end{equation}
This is a very restrictive Yukawa texture, giving non-vanishing mass to only one matter family. What happens for small deformations away from this wall into a stable chamber? 

To do this, one has to specify which of the two rank two sub-bundles in \eqref{so10split} 
destabilizes  $V$. Let us first  choose this to be ${\cal{F}}_{2}$. Then, $V$ can be constructed from the sequence
 \beq\label{extf2}
0 \to \cF_2 \to V \to {\cal F}_1 \to 0 
\eeq
and it is $C_2 \in H^1(X, \cF_2 \times {\cal F}_{1}^{*})$ that controls the movement away from the stability wall into the indecomposable gauge configuration. This can also be seen by inspecting the charges of the various fields in the $U(1)$ D-term associated with the wall, see \eref{dterm2}. Since we have chosen $\mu (\cF_2) <0$ in the stable region, it is the $C_2$ fields that must acquire a non-zero vev. Furthermore, the $(C_{1}C_2)^2$ term in \eqref{heirarchyeg} assures that  the vevs of $C_1$ must  vanish in the stable region. As a result, the two $C_{2}$ field dependent matter couplings in \eqref{heirarchyeg} ``grow back'' to contribute to Yukawa couplings near the stability wall. 
 It follows from the holomorphy analysis of subsection \ref{ancont} that one expects
 \begin{equation}
W_{\rm Yukawa} =h_1f_{2}^{2} +{\bf h_1f_{1}f_2+h_{1}f_{1}^{2} }
 \label{sun2}
 \end{equation}
{\it everywhere} in the interior of the stable chamber for this branch of the vacuum.
Note that if the K\"ahler moduli were stabilized {\it close to}, but not on, the stability wall, the four-dimensional $SO(10)$ theory would have one heavy family and a hierarchy for the remaining two generations controlled by powers of $\left<C_{2}\right>$.

Let us now consider the second branch where ${\cal{F}}_{1}$ is the destabilizing rank two sub-bundle. Then, $V$ can be constructed from the sequence
 \beq\label{extf2_again}
0 \to \cF_1 \to V \to {\cal F}_2 \to 0 
\eeq
and it is $C_1 \in H^1(X, \cF_1 \times {\cal F}_{2}^{*})$ that controls the movement away from the stability wall into the indecomposable gauge configuration. Since now $\mu ({\cal{F}}_{1}) < 0$, it follows from  $c_{1}(V)=0$ that $\mu ({\cal{F}}_{2})=-\mu ({\cal{F}}_{1}) > 0$. Hence, the FI term in 
$D^{U(1)}$, which is proportional to $\mu({\cal{F}}_{2})$, is positive and it is the $C_{1}$ fields that acquire a non-zero vev while the vevs of $C_{2}$ vanish. One must then conclude from 
\eqref{heirarchyeg} that {\it no} Yukawa couplings can ``grow back'' near the stability wall in this branch. It follows from the holomorphy analysis of subsection \ref{ancont} that one expects
\begin{equation}
W_{\rm Yukawa} = h_1f_{2}^{2} 
\label{sun3}
\end{equation}
{\it everywhere} in the interior of the stable chamber for this branch of the vacuum. Therefore, stability wall constraints can provide a natural way of obtaining a single heavy family in heterotic three family vacua.

\begin{table}[h]
\begin{center}
\begin{tabular}{|c|c|c|}
  \hline
  Representation & Field name & Cohomology \\ \hline
  $({\bf 1},{\bf 2},{\bf 2})_{2}$ & $C_1$ & $H^1(X,{\cal F}_1\otimes {\cal F}_2^*)$ \\ \hline
  $({\bf 1},{\bf 2},{\bf 2})_{-2}$ & $C_2$ & $H^1(X,{\cal F}_2\otimes {\cal F}_1^*)$ \\ \hline
 $({\bf 1},{\bf 3},{\bf 1})_{0}$ & $\phi_1$ & $H^1(X,{\cal F}_1 \otimes {\cal F}_1^*)$ \\ \hline
  $({\bf 1},{\bf 1},{\bf 3})_{0}$ & $\phi_2$ & $H^1(X,{\cal F}_2\otimes {\cal F}_2^*)$ \\ \hline
  $({\bf 16},{\bf 2},{\bf 1})_{1}$ & $f_1$ & $H^1(X,{\cal F}_1)$ \\ \hline
  $({\bf 16},{\bf 1},{\bf 2})_{-1}$ & $f_2$ & $H^1(X,{\cal F}_2)$ \\ \hline
  $(\overline{{\bf 16}},{\bf 2},{\bf 1})_{-1}$ & $\tilde{f}_1$ & $H^1(X,{\cal F}_1^*)$  \\ \hline
 $(\overline{{\bf 16}},{\bf 1},{\bf 2})_{1}$ & $\tilde{f}_2$ & $H^1(X,{\cal F}_2^*)$  \\ \hline
 $({\bf 10},{\bf 1},{\bf 1})_{2}$ & $h_1$ & $H^1(X,\wedge^2 {\cal F}_1)$ \\ \hline 
 $({\bf 10},{\bf 1},{\bf 1})_{-2}$ & $h_2$ & $H^1(X,\wedge^2 {\cal F}_2)$ \\ \hline 
 $({\bf 10},{\bf 2},{\bf 2})_{0}$ & $h_3$ & $H^1(X,{\cal F}_1 \otimes {\cal F}_2)$ \\ \hline 
  \end{tabular}
\mycaption{The spectrum of a {\it generic} $SU(4)$ bundle decomposing into two rank 2 bundles, $\cF_1 \oplus \cF_2$, on the stability wall. The resulting structure group is $S[U(2) \times U(2)]$.}
\label{tableso10sec}
\end{center}
\end{table}

\subsection{An Explicit Three Generation Model}\label{egso10}
In this subsection, we present an example of a three generation model with the stability wall structure described above and only one heavy family. 
To begin, consider a vector bundle over a simply connected Calabi-Yau threefold, $X$, which admits a fixed-point free, discrete automorphism $\Gamma=\mathbb{Z}_3 \times \mathbb{Z}_3$. This discrete symmetry allows one to construct a smooth quotient manifold $\hat{X}=X/(\mathbb{Z}_3 \times \mathbb{Z}_3)$ that is not simply connected\footnote{The first fundamental group of the quotient manifold is $\pi_1(\hat{X})=\mathbb{Z}_3\times \mathbb{Z}_3$.}. By choosing a vector bundle $V$ over the ``upstairs" manifold, $X$, which admits an equivariant structure under this symmetry, one can create a bundle $\hat{V}$ on the ``downstairs" threefold $\hat{X}$. 
This ``quotienting'' process is somewhat convoluted mathematically and, since it is not the central focus of this paper, we present here only the spectrum and properties of the final bundle $\hat{V}$ on $\hat{X}$. The derivation of this bundle in terms of its descent from the ``upstairs" theory, as well as relevant technical details, are given in Appendix \ref{equivariance}.

The Calabi-Yau threefold is taken to be a $\mathbb{Z}_3\times \mathbb{Z}_3$ quotient of the bi-cubic hypersurface in $\mathbb{P}^3 \times \mathbb{P}^3$ \cite{Anderson:2009mh},  
\beq\label{33q}
\hat{X}^{2,11}=X/\mathbb{Z}_{3} \times \mathbb{Z}_{3} \ , \quad X= \left[\begin{array}[c]{c}\mathbb{P}^2\\\mathbb{P}^2\end{array}
\left|\begin{array}[c]{ccc}3 \\3
\end{array}
\right.  \right]^{2,83}\; \ .
\eeq
Before giving a vector bundle on $\hat{X}$, we first describe how line bundles on $\hat{X}$ are related to those on $X$. Recall that $\cO(k,m)$ is the line bundle $\cO(k H_1 +mH_2)$, where $H_1,H_2$ are the restrictions of the hyperplanes of each $\mathbb{P}^2$ to $X$. The two-dimensional space of divisors of $X$ is spanned by $H_i$, $i=1,2$,  and we choose this as a convenient basis to describe divisors and the line bundles associated with them.
To describe a line bundle on $\hat{X}$, note that the basis of ample divisors on $\hat{X}$ is related to those on $X$ via the quotient map. Specifically, given the projection map $q: X \to \hat{X}$, a divisor $\hat{H}$ of $\hat{X}$ is related to some divisor $H$ on $X$ via $q^*(\hat{H})=H$. 
Using this, we choose a basis of divisors $\hat{H}_i$, $i=1,2$ on $\hat{X}$ to be related to $H_i$ via the pull-backs
\beq\label{divisors}
q^*(\hat{H}_1)=3H_1 \ , ~~~q^*(\hat{H}_2)=H_1 + H_2~.
\eeq
Using the divisor/line bundle correspondence, the basis of K\"ahler forms of $\hat{X}$ are then related to those on $X$ by 
$q^*(\hat{J}_1)=3J_1$ and $q^*(\hat{J}_2)=J_1+J_1$.

We define the rank four $SU(4)$ vector bundle on $\hat{X}$ via the exact sequence
\beq\label{downV}
0 \to \hat{\cF_{1}} \to \hat{V} \to \hat{\cF_{2}} \to 0 \ ,
\eeq
where $\hat{\cF_{1}},\hat{\cF_{2}}$ are rank two bundles constructed from
\beq
\ba
{c}
\label{downV2}
0 \to \hat{\cF_{1}} \to {\cal Q}_{1} \to \cO(2\hat{H}_2) \to 0 \ ,\\
0 \to \hat{\cF_{2}} \to {\cal Q}_{2} \to \cO(\hat{H}_2) \to 0 
\ea
\eeq
and ${\cal Q}_1,{\cal Q}_2$ are rank three bundles defined via their pull-backs
\beq
\label{downV3}
q^*({\cal Q}_{1})=\cO(0,2)^{\oplus 3}~~~~~q^*({\cal Q}_{2})=\cO(1,-1)^{\oplus 3}
\eeq 
 to sums of line bundles on $X$.
Since $c_1(\hat{\cF}_1)=(-2,4)$ and $c_1(\hat{\cF}_2)=(2,-4)$, then $c_1(\hat{V})=0$ and $\hat{V}$ in \eref{downV} defines an $SU(4)$ bundle over $\hat{X}$. The resulting four-dimensional theory has 
$SO(10)$ gauge symmetry. 
The matter spectrum of $\hat{V}$ is derived in Appendix \ref{equivariance} and given by
\beq
\ba
{l}
n_{16}=h^1(\hat{X},\hat{V})=h^1(\hat{X},\hat{\cF}_1)+h^1(\hat{X},\hat{\cF}_2)=2+1=3 \ , \\
n_{\bar{16}}=h^1(\hat{X}, \hat{V}^*)=0 \ , \label{sun4}\\
n_{10}=h^1(\hat{X},\wedge^2 \hat{V})=h^1(\hat{X},\wedge^2 \hat{\cF}_1)=2 \ ,
\ea
\eeq
which is very similar to \eqref{samplematter} in the preceding subsection, with the slight exception that there are two ${\bf 10}$'s.

The bundle, \eref{downV}, is not stable everywhere in the K\"ahler cone. By construction, $\hat{V}$ is destabilized by the bundle $\hat{\cF}_1$ in some region of K\"ahler moduli space. The region of stability is shown in Figure \ref{so10down}. This should be compared with the stability wall associated with the ``upstairs" bundle on $X$, presented in Figure \ref{so10up} of Appendix \ref{equivariance}\footnote{The stability wall structure of a bundle $\hat{V}$ on a quotient manifold is entirely determined by the stability structure of $V$ on $X$. Since only those sub-bundles of $V$ which are equivariant under the finite group action descend to sub-bundles of $\hat{V}$ on $\hat{X}$, the number of stability walls can at most decrease in going from $X$ to $\hat{X}$.}.
On the stability wall, $\hat{V}$ decomposes as $\hat{V} \rightarrow \hat{\cF}_1 \oplus \hat{\cF}_2$ and the structure group changes from $SU(4)$ to $S[U(2)\times U(2)]$. 
The $SO(10)$ gauge symmetry of the effective theory is then enhanced by an additional $U(1)$ symmetry to $SO(10)\times U(1)$. As a result, the ${\bf 16}$ and ${\bf 10}$ multiplets of $SO(10)$, as well as the bundle moduli, carry an additional charge. The decomposition of the cohomology under the enhanced symmetry group for this {\it explicit} example is presented in Table \ref{walldownstairs}. This is a subset of the generic spectrum of Table \ref{tableso10sec}. 
Note that, in addition to the matter multiplicities \eqref{sun4}, there are 9 $C_{1}$ type fields. However, no $C_{2}$ fields appear. Hence, this example describes the second branch of the generic vacuum discussed above. It follows that
\begin{equation}
W_{\rm Yukawa} = h_1f_{2}^{2} 
\label{sun5}
\end{equation}
{\it everywhere} in the interior of the stable chamber.
\begin{table}[h]
\begin{center}
\begin{tabular}{|c|c|c||c|}
  \hline
  Representation & Field name & Cohomology& Multiplicity \\ \hline
  $({\bf 1},{\bf 2},{\bf 2})_{2}$ & $\hat{C}_1$ & $H^1(\hat{X},\hat{{\cal F}}_1\otimes \hat{{\cal F}}_2^*)$& 9 \\ \hline
   $({\bf 1},{\bf 3},{\bf 1})_{0}$ & $\hat{\phi}_1$ & $H^1(\hat{X},\hat{{\cal F}}_1 \otimes \hat{{\cal F}}_1^*)$& 1 \\ \hline
  $({\bf 1},{\bf 1},{\bf 3})_{0}$ & $\hat{\phi}_2$ & $H^1(\hat{X},\hat{{\cal F}}_2\otimes \hat{{\cal F}}_2^*)$& 1 \\ \hline
  $({\bf 16},{\bf 2},{\bf 1})_{1}$ & $\hat{f}_1$ & $H^1(\hat{X},\hat{{\cal F}}_1)$& 2 \\ \hline
  $({\bf 16},{\bf 1},{\bf 2})_{-1}$ & $\hat{f}_2$ & $H^1(\hat{X},\hat{{\cal F}}_2)$& 1 \\ \hline
 $({\bf 10},{\bf 1},{\bf 1})_{2}$ & $\hat{h}_1$ & $H^1(\hat{X},\wedge^2 \hat{{\cal F}}_1)$&2 \\ \hline 
  \end{tabular}
\mycaption{The ``downstairs" field content of the {\it explicit} bundle decomposition $\hat{V}\rightarrow \hat{\cF}_1 \oplus \hat{\cF}_2$ defined by \eqref{downV}, \eqref{downV2} and \eqref{downV3}.}
\label{walldownstairs}
\end{center}
\end{table}
We conclude that this explicit vacuum naturally has one heavy family within the context of a realistic particle physics model.
 
 \begin{figure}[!ht]
  \centerline{\epsfxsize=5in\epsfbox{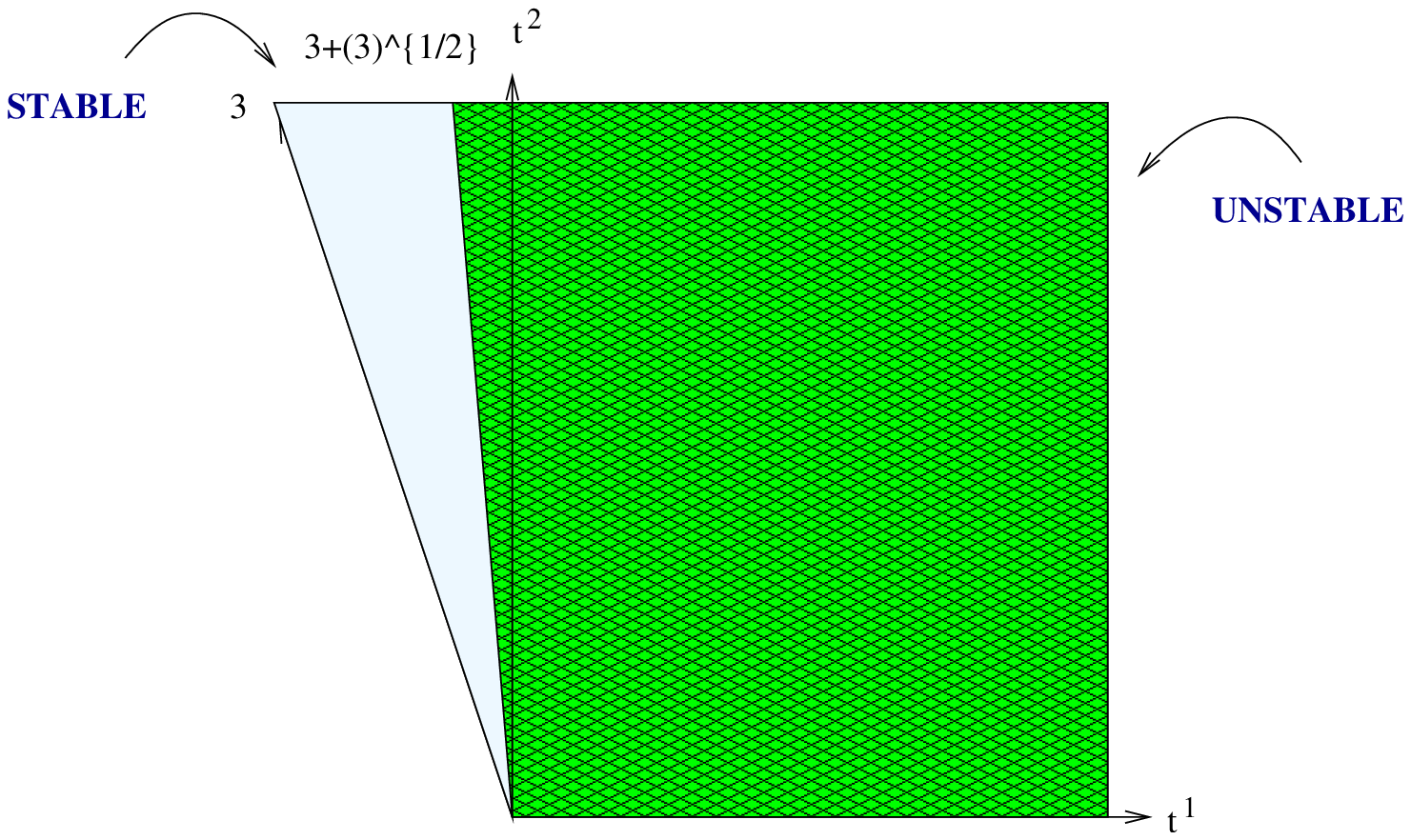}} \mycaption{
    The K\"ahler cone ($t^2>0$ and $t^2+3t^1>0$) and the regions of stability/instability for the ``downstairs" bundle $\hat{V}=V/(\mathbb{Z}_3\times\mathbb{Z}_3)$ on the quotient manifold $\hat{X}=X/(\mathbb{Z}_3\times\mathbb{Z}_3)$, defined respectively by \eqref{downV} and \eqref{33q}. At the line with slope $t^2/t^1=-3-\sqrt{3}$, $\hat{V}$ decomposes as $\hat{V} \rightarrow \hat{\cF_1} \oplus \hat{\cF_2}$ given in \eqref{downV2} and \eqref{downV3}.}
\label{so10down}
\end{figure}

\section{Constraints on Massive Vector-Like Pairs}\label{pairs}

Extended $U(1)$ gauge symmetry constrains the superpotential on and
near any stability wall and, by holomorphicity, in the interior of
each stable chamber in the K\"ahler cone.  So far, we have focused on
the implications of this for cubic matter interactions, that is,
Yukawa textures. However, the existence of stability walls constrains
all terms in the superpotential, not just Yukawa couplings. In this
section, we broaden our analysis to couplings involving vector-like
pairs of matter multiplets. We show that extended $U(1)$ symmetry can
forbid many, and sometimes all, such pairs from gaining superpotential
mass terms. This can have important implications for heterotic model
building. 

Generically, the zero-mode spectrum of a bundle on a stability wall arises from the cohomology of the sub-bundles into which it decomposes. In particular, matter can be in both a non-singlet representation and its conjugate representation of the low-energy gauge group. All such matter can occur on the stability wall, their multiplicity depending on the specific vacuum chosen. 
As one moves away from the wall into a stable chamber, the zero-mode spectrum can change. The Atiyah-Singer index theorem \cite{AG} requires that the chiral asymmetry of the matter representations be preserved. For example, for a stable $SU(3)$ bundle $V$ which decomposes into $V= {\cal{F}} \oplus {\cal{K}}$ on the stability wall, 
\begin{equation}
h^1(V) -h^1(V^*) = h^1({\cal F})-h^1({\cal F}^*) + h^1({\cal K})-h^1({\cal K}^*).
\label{t1}
\end{equation}
However, the actual number of matter  
representations need not stay the same. 
Specifically, as one moves away from the wall, certain $U(1)$ charged $C$ fields get a vev so as to preserve ${\cal{N}}=1$ supersymmetry. In principle, these can induce a non-vanishing mass for any vector-like pair of matter representations. As we have already seen, however, the extended $U(1)$ symmetry imposes serious constraints on cubic, and higher, matter couplings.  We expect there to be vector-like ``mass texture'' as well. As throughout this paper, we find it easiest to analyze vector-like pair masses within the context of explicit examples. 

\subsection{One Wall with One D-Term}

Let us first consider the class of vacua discussed in Subsection \ref{csection} and 
Section \ref{yukawawall}. In this case, $h^{1,1}(X)=2$ and   $V$ is an $SU(3)$ bundle which decomposes at a single stability wall into $V={\cal{F}} \oplus {\cal{K}}$, where ${\cal{F}}$ and ${\cal{K}}$ have rank one and two respectively. The generic spectrum on the wall arises as the product cohomologies of 
${\cal{F}}$ and ${\cal{K}}$, and is labeled by representations of the extended $E_{6} \times U(1)$ four-dimensional gauge group.  This is presented in Table \ref{table2}. The most general gauge invariant superpotential involving terms {\it cubic} in the $F$'s was given in \eqref{Wgeneral}. We now extend this to include all relevant terms involving ${\bf 27\cdot {\overline{27}}}$ {\it vector-like pairs} of  matter multiplets. The result is 
\begin{equation}
W=\dots+ C_{1}F_{1}{\bar{F}}_{2}+C_{2}F_{2}{\bar{F}}_{1}+C_{1}C_{2}F_{1}{\bar{F}}_{1}
+C_{1}C_{2}F_{2}{\bar{F}}_{2},
\label{t2}
\end{equation}
where terms are shown in order of increasing dimension and we have suppressed all parameters and indices. Note that no quadratic terms appear, since, on the wall, all matter fields are zero-modes. Finally, each term can be multiplied by any positive power of $C_{1}C_{2}$. Such terms do not change the subsequent analysis and, in the interest of brevity, we ignore them.

As discussed previously, {\it on the stability wall} the requirement of ${\cal{N}}=1$ supersymmetry and vanishing cosmological constant constrains $\left<C_{1}\right> =\left<C_{2}\right>=0$. It follows from \eqref{t2} that 
\begin{equation}
W^{\rm wall}_{\rm vec-like \ pairs}=0 \ ,
\label{t3}
\end{equation}
consistent with the fact that $F_{1}$, $F_{2}$ and ${\bar{F}}_{1}$, ${\bar{F}}_{2}$
are all zero-modes on the wall. What happens as we move into the interior a stable region? As discussed in Section \ref{csection}, there are two stable branches of moduli space. These are specified by choosing either $\left<C_{1}\right> =0, \left<C_{2}\right>\neq 0$, corresponding to 
$\mu ({\cal{F}}) < 0$, or $\left<C_{1}\right> \neq 0, \left<C_{2}\right> = 0$, corresponding to 
$\mu ({\cal{K}}) < 0$. Consider the first branch. In this case, it follows from \eqref{t2} and the holomorphicity of the superpotential that {\it everywhere} in this chamber of K\"ahler moduli space
\begin{equation}
W_{\rm vec-like \ pairs}= {\bf F_{2}{\bar{F}}_{1}} \ .
\label{t4}
\end{equation}
Note that the non-zero $C_{2}$ vevs have allowed some vector-like mass terms missing in \eqref{t3} to ``grow back''. These are expressed in boldface,  as were Yukawa couplings that regrew away the wall.
We conclude that in the interior of the stable chamber specified by 
$\left<C_{1}\right> =0, \left<C_{2}\right>\neq 0$, 
superfields $F_{2}$ and ${\bar{F}}_{1}$ appear in non-vanishing mass terms. However, the extended $U(1)$ gauge symmetry on the stability wall {\it forbids} vector-like masses for $F_{1}$ and ${\bar{F}}_{2}$ from developing.
Now consider the second branch. In this case, it follows from \eqref{t2} and holomorphicity that 
{\it everywhere} in this stable chamber
\begin{equation}
W_{\rm vec-like \ pairs}= {\bf F_{1}{\bar{F}}_{2}} \ .
\label{t5}
\end{equation}
Hence, in the interior of the stable chamber specified by 
$\left<C_{1}\right> \neq0, \left<C_{2}\right>= 0$, the extended $U(1)$ gauge symmetry on the stability wall, while allowing superfields $F_{1}$ and ${\bar{F}}_{2}$ to appear in mass terms, {\it forbids} vector-like masses for $F_{2}$ and ${\bar{F}}_{1}$. 

This is a clear example where the stable chambers next to a stability wall exhibit non-trivial {\it vector-like mass textures}; allowing some mass terms while forbidding others.

\subsection{One Wall with Two D-Terms}

We now move on to consider the class of vacua discussed in Section
\ref{onewall2D}. In this case, $h^{1,1}(X)=2$ and $V$ is an $SU(3)$
bundle which decomposes at a single stability wall into $V=l_{1}
\oplus l_{2} \oplus l_{3}$, where $l_{i}$, $i=1,2,3$ are line
bundles. The generic spectrum on the wall arises as the product
cohomologies of $l_{1}$, $l_{2}$, $l_{3}$ and is labeled by
representations of the extended $E_{6} \times U(1) \times U(1)$
four-dimensional gauge group.  This is presented in Table
\ref{tabledoubled}. The most general gauge invariant superpotential
involving {\it cubic} couplings in the $F$'s was given in \eqref{2Dw}. We
now extend this result to include all relevant terms involving ${\bf
  27\cdot {\overline{27}}}$ {\it vector-like pairs} of matter
multiplets. The result is
\begin{eqnarray}
W= \dots &+&C_{1}f_{1}{\tilde{f}}_{2}+C_{2}f_{1}{\tilde{f}}_{3}+{\tilde{C}}_{1}f_{2}{\tilde{f}}_{1}+{\tilde{C}}_{3}f_{2}{\tilde{f}}_{3}+ {\tilde{C}}_{2}f_{3}{\tilde{f}}_{1}+{C}_{3}f_{3}{\tilde{f}}_{2}  \nonumber\\
&+&\left( C_{1}{\tilde{C}}_{1} +C_{2}{\tilde{C}}_{2} + C_{3}{\tilde{C}}_{3}\right) \left( f_{1} {\tilde{f}}_{1} + f_{2} {\tilde{f}}_{2}+ f_{3} {\tilde{f}}_{3}\right) \label{t6} \\
&+&C _{1}{\tilde{C}}_{2}f_{3}{\tilde{f}}_{2}+C _{1}{\tilde{C}}_{3}f_{1}{\tilde{f}}_{3}+{\tilde{C}} _{1}C_{2}f_{2}{\tilde{f}}_{3}+{\tilde{C}} _{1}C_{3}f_{3}{\tilde{f}}_{1}+C _{2}C_{3}f_{1}{\tilde{f}}_{2}+{\tilde{C}} _{2}{\tilde{C}}_{3}f_{2}{\tilde{f}}_{1} \nonumber 
\end{eqnarray}
No quadratic terms appear since all superfields are zero-modes on the wall. We have only indicated terms involving at most two different $C$ fields. Vevs of the product of three or more different $C$ fields must necessarily vanish in any branch. Finally, each term can be multiplied by any positive integer power of neutral combinations of $C$ fields. Such terms do not change the subsequent analysis.

{\it On the stability wall}, the requirement of supersymmetry and vanishing cosmological constant constrains the vevs of each $C$ field to vanish. Hence,
\begin{equation}
W^{\rm wall}_{\rm vec-like \ pairs}=0 \ ,
\label{t7}
\end{equation}
consistent with the fact that all $f$ and $\tilde{f}$ matter fields are zero-modes on the wall. What happens as we move into a stable region? As discussed in Section \ref{onewall2D}, there are {\it six} stable branches of the moduli space. Each branch is specified by a different pair $(\left<C_{i}\right>,\left<C_{j}\right>)$, ($\left<C_{i}\right>,\left<{\tilde{C}}_{j}\right>)$ or 
$(\left<{\tilde{C}}_{i}\right>,\left<{\tilde{C}}_{j}\right>)$ being non-vanishing, with all remaining vevs zero. To be specific, let us choose the branch defined by $\left< \tilde{C}_{2}\right> \neq 0$,
$\left<C_{3}\right> \neq 0$. It then follows from \eqref{t3} and holomorphicity that in the interior of this branch of K\"ahler moduli space
\begin{equation}
W_{\rm vec-like \ pairs}={\bf  f_{3}{\tilde{f}}_{1}+f_{3}{\tilde{f}}_{2}} \ . 
\label{t8}
\end{equation}
Note that the non-zero ${\tilde{C}}_{2}$, $C_{3}$ vevs have allowed some vector-like mass terms missing in \eqref{t7} to ``grow back''. Therefore, in the interior of the stable chamber specified by 
$\left<{\tilde{C}}_{2}\right> \neq 0, \left<C_{3}\right>\neq 0$, 
matter multiplets $f_{3}$, ${\tilde{f}}_{1}$ and ${\tilde{f}}_{2}$ appear in non-vanishing mass terms. However, the two extended $U(1)$ gauge symmetries on the stability wall {\it forbid} vector-like masses for $\tilde{f}_{3}$,  $f_{1}$ and $f_{2}$ from developing.

We conclude that the extended $U(1)$ gauge symmetries on stability walls in the K\"ahler cone can lead to restrictive {\it vector-like mass textures}. Generically, these textures can disallow some vector-like pairs from having a superpotential mass term, a restriction of consequence for phenomenology. Hence, when building realistic smooth heterotic models, it is essential to include all stability walls and their associated constraints in the analysis. This makes theories with only chiral matter appear much more attractive from this perspective.

\section{Conclusions}\label{conclusions}
In previous work \cite{Sharpe:1998zu,Anderson:2009sw,Anderson:2009nt},
``stability walls", that is, boundaries separating regions in K\"ahler moduli
space where a non-Abelian internal gauge bundle either preserves or breaks
supersymmetry, were explored. The four-dimensional effective theories
valid near such boundaries provide us with an explicit low-energy
description of the supersymmetry breaking associated with vector bundle
slope stability. The central feature of a stability wall is that, near
such a locus in moduli space, the internal gauge bundle decomposes
into a direct sum and, as a result, the four-dimensional effective
theory is enhanced by at least one Green-Schwarz anomalous $U(1)$
symmetry.

In this paper, we have used this effective theory to investigate the structure and properties of heterotic theories with
stability induced sub-structure in their K\"ahler cones. Specifically,
we have used the theory near the stability wall, with its enhanced
$U(1)$ symmetries, to constrain the form of the ${\cal N}=1$
superpotential $W$. Using the fact that the superpotential is a
holomorphic function, it is possible to extend these constraints
throughout the entire moduli space. As a result, deep into the stable
regions of the K\"ahler cone, where supersymmetric heterotic
compactifications are normally considered, strong constraints on the
superpotential still persist. Without knowledge of the global
supersymmetric properties of the vector bundle (that is, a full
understanding of its slope stability), these important textures would
be inexplicable or, more seriously, go unnoticed if
the Yukawa couplings were not explicitly computed. We would like to point out 
that some of the couplings that are disallowed in the perturbative
textures discussed in this paper may well be reintroduced by non-perturbative
effects, such as membrane instantons \cite{Lima:2001jc,Braun:2007tp}. Such couplings would be
hierarchically smaller than those present
perturbatively. This interesting possibility, which also is strongly
constrained by the additional Abelian symmetries on the 
stability walls, will be addressed in a future publication.

We stress again that the existence of stability walls, and
their consequences, are the generic situation for a heterotic
compactification. In most cases, vector bundles that are slope stable somewhere in moduli space are not slope stable
for all polarizations.  Hence, the constraints described in this paper must
be considered to have a full understanding of the effective
theory. Indeed, all three of the main methods of bundle construction in the heterotic literature-- monad bundles
\cite{Anderson:2007nc,Anderson:2008uw,Anderson:2008ex,He:2009wi,Anderson:2009mh},
bundles defined by extension \cite{AG,burt} and the 
spectral cover construction
\cite{Friedman:1997yq,Friedman:1997ih,Donagi:2000zs} - typically
exhibit stability walls. Thus, from the point of view of model
building and string phenomenology, the textures and constraints on
Yukawa couplings and vector - like masses discussed in this paper are generically present. They must be taken into account in any attempt to build physically realistic models.

\section*{Acknowledgements}
J.~Gray is supported by STFC and would like to thank the Department of
Physics and Astronomy at the University of Pennsylvania for
hospitality while some of this work was being completed. L.~Anderson
and B.~A.~Ovrut are supported in part by the DOE under contract No.
DE-AC02-76-ER-03071 and by NSF RTG Grant DMS-0636606.
 \appendix
 
 \section{All Textures from a Single D-term Stability Wall }

In this Appendix, we present all Yukawa textures
that can result from holomorphic vector bundles with a {\it single} stability wall where the bundle splits into a direct sum of {\it two factors}. These are an important sub-class of
Yukawa textures that can appear naturally within the context of 
heterotic string and M-theory.

\subsection{An $SU(3)$ Bundle with a Stability Wall and One D-Term}

We begin with a compactification of heterotic theory on a Calabi-Yau threefold with a rank {\it three} holomorphic vector bundle $V$. For any K\"ahler form in the stable chamber, the structure group is an indecomposable $SU(3)$ leading to an $E_6$ gauge group in the low-energy theory.
At the stability wall, where the bundle splits into two parts, an
$SU(3)$ bundle necessarily breaks into a rank 2 and a rank 1 piece, which
we will denote by ${\cal F}$ and ${\cal K}$ respectively.  That is, \bea V =
{\cal F} \oplus {\cal K} \ . \eea
The structure group of this bundle is $S[U(2) \times U(1)] \cong SU(2) \times U(1)$, leading to an enhanced $E_{6} \times U(1)$ gauge group in the effective theory. It follows that
the relevant decomposition of the $\bf{248}$ of $E_{8}$ is  \bea
E_8 &\supset& E_6 \times SU(2) \times U(1) \\
{\bf 248} &=& ({\bf 1},{\bf 1})_0 + ({\bf 1},{\bf 2})_3 + ({\bf
  1},{\bf 2})_{-3} + ({\bf 1},{\bf 3})_0 + ({\bf 78},{\bf 1})_0 \\
\nonumber &&+ ({\bf 27},{\bf 1})_{-2} + ({\bf 27},{\bf 2})_1 +
(\overline{{\bf 27}},{\bf 1})_2 + (\overline{{\bf 27}},{\bf 2})_{-1} \ .
\eea This decomposition indicates which representations of $E_{6} \times U(1)$ can possibly appear as fields in the  four-dimensional effective theory. To find out how many of each multiplet is actually present, one must calculate the dimension of the cohomology groups indicated in
Table \ref{tableA1}.
\begin{table}[h]
\begin{center}
\begin{tabular}{|c|c|c|}
  \hline
  Representation & Field name & Cohomology \\ \hline
  $({\bf 1},{\bf 2})_{3}$ & $C_1$ & $h^1(X,{\cal F}^*\otimes {\cal K})$ \\ \hline
  $({\bf 1},{\bf 2})_{-3}$ & $C_2$ & $h^1(X,{\cal F} \otimes {\cal K}^*)$  \\ \hline
  $({\bf 1},{\bf 3})_{0}$ & $\phi$ & $h^1(X,{\cal F}^* \otimes {\cal F})$  \\ \hline
  $({\bf 27},{\bf 1})_{-2}$ & $f_1$ & $h^1(X,{\cal K})$ \\ \hline
  $({\bf 27},{\bf 2})_{1}$ & $f_2$ & $h^1(X,{\cal F})$  \\ \hline
  $(\overline{{\bf 27}},{\bf 1})_{2}$ & $\tilde{f}_1$ & $h^1(X,{\cal K}^*)$  \\ \hline  $(\overline{{\bf 27}},{\bf 2})_{-1}$ & $\tilde{f}_2$ & $h^1(X,{\cal 
F}^*)$  \\ \hline
  \end{tabular}
\mycaption{$SU(3)$ one D-term.}
\label{tableA1}
\end{center}
\end{table}
As discussed in Section 3, only one of the two fields $C_1$,$C_2$ can get a vev. If $\left< C_1 \right> \neq 0$, then the allowed Yukawa couplings are 
 \bea W_{\textnormal{Yukawa}} = f_1 f_2^2 + \tilde{f}_1 \tilde{f}_2^2 + {\bf f_1^3 + f_1^2
f_2 + \tilde{f}_2^3} \ . \label{buddy1} \eea
As discussed in the text, we suppress the arbitrary coefficients in front of each term for simplicity. Terms allowed in the dimension three (in superfields) superpotential on the stability wall are shown in standard type. Yukawa terms that originate as higher dimensional operators involving powers of $C_{1}$, which are ``grown back'' upon re-entering the
interior supersymmetric region where $<C_{1}> \neq 0$, are indicated in boldface.
On the other hand, if $\left< C_2 \right> \neq 0$, then we find \bea
W_{\textnormal{Yukawa}} = f_1 f_2^2 + \tilde{f}_1 \tilde{f}_2^2 + {\bf  \tilde{f}_1^3
+\tilde{f}_1^2 \tilde{f}_2 + f_2^3}  \;.  \eea 
Note that on the stability wall the $U(1)$ charges strongly restrict the allowed Yukawa couplings. As one moves away from the stability wall into the stable chamber, a number of previously disallowed couplings ``grow back''. However, not all terms allowed by the $E_{6}$ symmetry can reappear. For example, an
${\bf  \tilde{f}_1^3}$ term can never be generated in superpotential \eqref{buddy1}. 

\subsection{An $SU(4)$ Bundle with a Stability Wall and One D-Term}

Now consider a compactification of heterotic theory on a Calabi-Yau threefold with a rank {\it four} holomorphic vector bundle $V$. For any K\"ahler form in the stable chamber, the structure group is an indecomposable $SU(4)$ leading to an $SO(10)$ gauge group in the low-energy theory.
There are now two ways in which the bundle can split at a
stability wall. We treat each case in turn. We emphasize that both cases can be realized by rank four bundles with a single stability wall.

\subsubsection{Case $1$ }
The first of the two cases corresponds to the bundle splitting into a
rank 3 and a rank 1 piece, which we shall denote by ${\cal F}$ and
${\cal K}$ respectively.  That is, \bea V = {\cal F} \oplus {\cal K} \ . \eea
The structure group of this bundle is $SU(3) \times U(1)$, leading to an 
enhanced $SO(10) \times U(1)$ 
gauge group in the effective theory. The relevant group theory here is
\bea
E_8 &\supset& SO(10) \times SU(3) \times U(1) \\
{\bf 248} &=& ({\bf 1},{\bf 1})_0 + ({\bf 1},{\bf 3})_{-4} + ({\bf
  1},\overline{{\bf 3}})_4 + ({\bf 1},{\bf 8})_0 + ({\bf 45},{\bf
  1})_0 + ({\bf 16},{\bf 1})_3 + ({\bf 16},{\bf 3})_{-1} \\ \nonumber &&+
(\overline{{\bf 16}},{\bf 1})_{-3} + (\overline{{\bf
    16}},\overline{{\bf 3}})_1 + ({\bf 10},{\bf 3})_2 + ({\bf
  10},\overline{{\bf 3}})_{-2}\;.\eea
This decomposition indicates which representations of $SO(10) \times U(1)$ can possibly appear as fields in the  four-dimensional effective theory. To find out how many of each multiplet is actually present, one must calculate the dimension of the cohomology groups indicated in
Table \ref{tableA2}.
\begin{table}[h]
\begin{center}
\begin{tabular}{|c|c|c|}
  \hline
  Representation & Field name & Cohomology \\ \hline
  $({\bf 1},{\bf 3})_{-4}$ & $C_1$ & $h^1(X,{\cal F}\otimes {\cal K}^*)$ \\ \hline
  $({\bf 1},\overline{{\bf 3}})_{4}$ & $C_2$ & $h^1(X,{\cal F}^* \otimes {\cal K})$  \\ \hline
  $({\bf 1},{\bf 8})_{0}$ & $\phi$ & $h^1(X,{\cal F}^* \otimes {\cal F})$  \\ \hline
  $({\bf 16},{\bf 1})_{3}$ & $f_1$ & $h^1(X,{\cal K})$ \\ \hline
  $({\bf 16},{\bf 3})_{-1}$ & $f_2$ & $h^1(X,{\cal F})$ \\ \hline
  $(\overline{{\bf 16}},{\bf 1})_{-3}$ & $\tilde{f}_1$ & $h^1(X,{\cal K}^*)$  \\ \hline  $(\overline{{\bf 16}},\overline{{\bf 3}})_{1}$ & $\tilde{f}_2$ & $h^1(X,{\cal 
    F}^*)$  \\ \hline
  $({\bf 10},{\bf 3})_{2}$ & $h_1$ & $h^1(X,{\cal F} \otimes {\cal K})$ \\ \hline 
  $({\bf 10},\overline{{\bf 3}})_{-2}$ & $h_2$ & $h^1(X,\wedge^2 {\cal F})$ \\ \hline
  \end{tabular}
\mycaption{$SU(4)$ one D-term case 1.}
\label{tableA2}
\end{center}
\end{table}
As discussed previously, only one of the two $C_1$,$C_2$ fields can have a non-zero vev.  
If $\left< C_1 \right> \neq 0$, then the allowed Yukawa couplings are  
\bea W_{\textnormal{Yukawa}}=
h_1 f_2^{2} + h_1\tilde{f}_1 \tilde{f}_2+ h_2 \tilde{f}_2^2 +{\bf h_1 f_1^2 + h_1 f_1 f_2 + h_2 f_1^2 + h_2 f_2 f_2 +  h_1 \tilde{f}_2^2} \; . \eea
On the other hand, if $\left< C_2 \right> \neq 0$, then we find
\bea W_{\textnormal{Yukawa}} = h_1
f_2^2 +h_1 \tilde{f}_1
\tilde{f}_2 + h_2 \tilde{f}_2^2+ {\bf h_2 f_1 f_2 +h_2 f_2^2 + h_2 \tilde{f}_1 \tilde{f}_2 + h_1 \tilde{f}_1^2 + h_2 \tilde{f}_1^2}   \;.\eea
\subsubsection{Case $2$}\label{su4case2}
The second $SU(4)$ case corresponds to the rank 4 bundle
splitting into two rank 2 pieces denoted by ${\cal F}_1$ and  ${\cal F}_2$. That is,
 \bea V = {\cal F}_1 \oplus {\cal
  F}_2 \ . \eea 
Under this splitting the structure group breaks into $SU(2) \times SU(2) \times U(1)$, leading to an enhanced $SO(10) \times U(1)$ gauge group in the effective theory. The relevant branchings are
\bea E_8 &\supset& SO(10) \times SU(2) \times SU(2) \times U(1) \\
{\bf 248} &=& ({\bf 1},{\bf 1},{\bf 1})_0 + ({\bf 1},{\bf 3},{\bf
  1})_0 +({\bf 1},{\bf 1},{\bf 3})_0 + ({\bf 1}, {\bf 2},{\bf 2})_2 +
({\bf 1},{\bf 2},{\bf 2}))_{-2} + ({\bf 45},{\bf 1},{\bf 1})_0  \\
&+&({\bf 16},{\bf 2},{\bf 1})_1 + ({\bf 16},{\bf 1},{\bf
  2})_{-1} + (\overline{{\bf 16}},{\bf 2},{\bf 1})_{-1} +
(\overline{{\bf 16}},{\bf 1},{\bf 2})_1 + ({\bf 10},{\bf 1},{\bf 1})_2
+ ({\bf 10},{\bf 1},{\bf 1})_{-2} \nonumber \\
&+&({\bf 10},{\bf 2},{\bf 2})_0 \ .\nonumber \eea
The cohomology associated with each field is given in Table \ref{tableA3}. 
As before, only one of the $C_1$, $C_2$ fields can get a vev.
\begin{table}[h]
\begin{center}
\begin{tabular}{|c|c|c|}
  \hline
  Representation & Field name & Cohomology \\ \hline
  $({\bf 1},{\bf 2},{\bf 2})_{2}$ & $C_1$ & $h^1(X,{\cal F}_1\otimes {\cal F}_2^*)$ \\ \hline
  $({\bf 1},{\bf 2},{\bf 2})_{-2}$ & $C_2$ & $h^1(X,{\cal F}_2\otimes {\cal F}_1^*)$ \\ \hline
 $({\bf 1},{\bf 3},{\bf 1})_{0}$ & $\phi_1$ & $h^1(X,{\cal F}_1 \otimes {\cal F}_1^*)$ \\ \hline
  $({\bf 1},{\bf 1},{\bf 3})_{0}$ & $\phi_2$ & $h^1(X,{\cal F}_2\otimes {\cal F}_2^*)$ \\ \hline
  $({\bf 16},{\bf 2},{\bf 1})_{1}$ & $f_1$ & $h^1(X,{\cal F}_1)$ \\ \hline
  $({\bf 16},{\bf 1},{\bf 2})_{-1}$ & $f_2$ & $h^1(X,{\cal F}_2)$ \\ \hline
  $(\overline{{\bf 16}},{\bf 2},{\bf 1})_{-1}$ & $\tilde{f}_1$ & $h^1(X,{\cal F}_1^*)$  \\ \hline
 $(\overline{{\bf 16}},{\bf 1},{\bf 2})_{1}$ & $\tilde{f}_2$ & $h^1(X,{\cal F}_2^*)$  \\ \hline
 $({\bf 10},{\bf 1},{\bf 1})_{2}$ & $h_1$ & $h^1(X,\wedge^2 {\cal F}_1)$ \\ \hline 
 $({\bf 10},{\bf 1},{\bf 1})_{-2}$ & $h_2$ & $h^1(X,\wedge^2 {\cal F}_2)$ \\ \hline 
 $({\bf 10},{\bf 2},{\bf 2})_{0}$ & $h_3$ & $h^1(X,{\cal F}_1 \otimes {\cal F}_2)$ \\ \hline 
  \end{tabular}
\mycaption{$SU(4)$ one D-term case 2.}
\label{tableA3}
\end{center}
\end{table}
Which vev is non-zero determines the structure of the Yukawa
couplings. For $<C_{1}> \neq 0$, we find that  
\bea \label{egtext}
W_{\textnormal{Yukawa}} &=& h_1 f_2^2 +
h_2 f_1^2 + h_3 f_1 f_2 +h_1
\tilde{f}_1^2 + h_2 \tilde{f}_2^2 +h_3 \tilde{f}_1
\tilde{f}_2\\
\nonumber &+&{\bf h_2 f_1 f_2 + h_2 f_2^2 + h_3 f_2^2+ h_2 \tilde{f}_1^2 + h_2 \tilde{f}_1
\tilde{f}_2 +  h_3 \tilde{f}_1^2}  \;,\eea whereas for $<C_{2}> \neq 0$ the following texture
appears 
\bea W_{\textnormal{Yukawa}} &=& h_1 f_2^2 + h_2 f_1^2 +h_3 f_1 f_2+h_1 \tilde{f}_1^2 +h_2
\tilde{f}_2^2 + h_3 \tilde{f}_1 \tilde{f}_2  \\ \nonumber
&+& {\bf h_1 f_1 f_2 + h_1f_1^2 + h_3 f_1^2 +h_1 \tilde{f}_1 \tilde{f}_2 + h_1 \tilde{f}_2^2 + h_3 \tilde{f}_2^2}
\;. \eea

\subsection{An $SU(5)$ Bundle with a Stability Wall and One D-Term}

Consider compactification of heterotic theory on a Calabi-Yau threefold with a rank {\it five} holomorphic vector bundle $V$. For any K\"ahler form in the stable chamber, the structure group is an indecomposable $SU(5)$ leading to an $SU(5)$ gauge group in the low-energy theory.
As in the $SU(4)$ case, there are two ways in which such bundles can split at a stability
wall. We treat these sequentially.
\subsubsection{Case $1$}
First consider the case where the bundle splits into a rank 4 and
a rank 1 piece, denote by ${\cal F}$ and ${\cal K}$
respectively. That is,  \bea V = {\cal F} \oplus {\cal K} \ . \eea 
The structure group of this bundle is $SU(4) \times U(1)$, leading to an 
enhanced $SU(5) \times U(1)$ 
gauge group in the effective theory. The relevant branching of the ${\bf 248}$ representation here is
 \bea
E_8 &\supset& SU(5) \times SU(4) \times U(1) \\
{\bf 248} &=& ({\bf 24},{\bf 1})_0 +({\bf 1},{\bf 1})_0 + ({\bf
  1},{\bf 4})_{-5}+({\bf 1},\overline{{\bf 4}})_5+ ({\bf 1},{\bf
  15})_0 + ({\bf 10},{\bf 1})_4+({\bf 10},{\bf 4})_{-1} \\ \nonumber
&&+ (\overline{{\bf 10}},{\bf 1})_{-4}+(\overline{{\bf
    10}},\overline{{\bf 4}})_1 + ({\bf 5},\overline{{\bf
    4}})_{-3}+({\bf 5},\overline{{\bf 6}})_{2} + (\overline{{\bf
    5}},{\bf 4})_3+ (\overline{{\bf 5}},{\bf 6})_{-2} \;.\eea The
multiplicities of the matter multiplets, in terms of cohomologies of
${\cal F}$ and ${\cal K}$, may be found in Table \ref{tableA4}.
\begin{table}[h]
\begin{center}
\begin{tabular}{|c|c|c|}
  \hline
  Representation & Field name & Cohomology \\ \hline
  $({\bf 1},{\bf 4})_{-5}$ & $C_1$ & $h^1(X,{\cal F}\otimes {\cal K}^*)$ \\ \hline
  $({\bf 1},\overline{{\bf 4}})_{5}$ & $C_2$ & $h^1(X,{\cal F}^* \otimes {\cal K})$  \\ \hline
  $({\bf 1},{\bf 15})_{0}$ & $\phi$ & $h^1(X,{\cal F}^* \otimes {\cal F})$  \\ \hline
  $({\bf 10},{\bf 1})_{4}$ & $f_1$ & $h^1(X,{\cal K})$ \\ \hline
  $({\bf 10},{\bf 4})_{-1}$ & $f_2$ & $h^1(X,{\cal F})$ \\ \hline
  $(\overline{{\bf 10}},{\bf 1})_{-4}$ & $\tilde{f}_1$ & $h^1(X,{\cal K}^*)$  \\ \hline  $(\overline{{\bf 10}},\overline{{\bf 4}})_{1}$ & $\tilde{f}_2$ & $h^1(X,{\cal 
    F}^*)$  \\ \hline
  $({\bf 5},\overline{{\bf 4}})_{-3}$ & $h_1$ & $h^1(X,{\cal F}^* \otimes {\cal K}^*)$ \\ \hline 
  $({\bf 5},\overline{{\bf 6}})_{2}$ & $h_2$ & $h^1(X,\wedge^2 {\cal F}^*)$ \\ \hline
  $(\overline{{\bf 5}},{\bf 4})_{3}$ & $\tilde{h}_1$ & $h^1(X,{\cal F} \otimes {\cal K})$ \\ \hline 
  $(\overline{{\bf 5}},{\bf 6})_{-2}$ & $\tilde{h}_2$ & $h^1(X,\wedge^2 {\cal F} )$ \\ \hline
  \end{tabular}
\mycaption{$SU(5)$ one D-term case 1.}
\label{tableA4}
\end{center}
\end{table}
As in the previous cases, there are two possible textures. 
For $<C_{1}> \neq 0$, we find that  
\bea W_{\textnormal{Yukawa}} &=&
h_1 f_1 f_2 + + h_2 f_2^2+{\bf h_1 f_1^2 + h_2 f_1^2 + h_2 f_1 f_2}  \\
\nonumber && + \tilde{h}_1 \tilde{f}_1
\tilde{f}_2 +  \tilde{h}_2 \tilde{f}_2^2+ {\bf \tilde{h}_1 \tilde{f}_2^2}  \\
\nonumber && +f_1 \tilde{h}_2^2+ f_2 \tilde{h}_1 \tilde{h}_2+{\bf f_1 \tilde{h}_1^2 + 
f_1 \tilde{h}_1 \tilde{h}_2 + f_2 \tilde{h}_1^2}  \\
\nonumber && + \tilde{f}_1 h_2^{2}+ \tilde{f}_2 h_1 h_2 +{\bf  \tilde{f}_2
h_2^2} \;,\eea 
where we have grouped all of the ${\bf 5 \cdot10\cdot 10}$, ${\bf
  \overline{5} \cdot\overline{10} \cdot\overline{10}}$, ${\bf 10 \cdot\overline{5}
 \cdot \overline{5}}$ and ${\bf \overline{10} \cdot 5 \cdot5}$ couplings together
on different lines. For $<C_{2}> \neq 0$, the following texture
appears 
\bea W_{\textnormal{Yukawa}} &=& h_1
f_1 f_2 + h_2 f_2 ^2 +{\bf h_1 f_2^2 }\\ \nonumber &&+ \tilde{h}_1 \tilde{f}_1 \tilde{f}_2 + \tilde{h}_2
\tilde{f}_2^2 +{\bf \tilde{h}_1
\tilde{f}_1^2 + \tilde{h}_2
\tilde{f}_1^2 + \tilde{h}_2 \tilde{f}_1 \tilde{f}_2}  \\ \nonumber && + f_1 \tilde{h}_2^2 + f_2 \tilde{h}_1
\tilde{h}_2 + {\bf f_2 \tilde{h}_2^2} \\ \nonumber && +\tilde{f}_1 h_2^2+\tilde{f}_2 h_1 h_2 + {\bf \tilde{f}_1 h_1^{2}+\tilde{f}_1 h_1 h_2  + \tilde{f}_2 h_1^2} \;.\eea

\subsubsection{Case $2$}
Second, a rank 5 bundle can split into a rank 3 and a rank 2 piece,
${\cal G}$ and ${\cal F}$ respectively, at the stability wall. That is,
 \bea V = {\cal G} \oplus {\cal F} \ . \eea 
The structure group of this bundle is $SU(2) \times SU(3) \times U(1)$, leading to an 
enhanced $SU(5) \times U(1)$ 
gauge group in the effective theory. The relevant branching of the ${\bf 248}$ representation is given by
\bea
E_8 &\supset& SU(5) \times SU(2) \times SU(3) \times U(1) \\
{\bf 248} &=& ({\bf 24},{\bf 1},{\bf 1})_0+({\bf 1},{\bf 1},{\bf
  1})_0+({\bf 1}, {\bf 3},{\bf 1})_0+({\bf 1},{\bf 2},{\bf 3})_{-5} +
({\bf 1},{\bf 2},\overline{{\bf 3}})_5 +({\bf 1},{\bf 1},{\bf
  8})_0 \\
\nonumber &&+({\bf 10},{\bf 2},{\bf 1})_3+ ({\bf 10},{\bf 1},{\bf
  3})_{-2} + (\overline{{\bf 10}},{\bf 2},{\bf
  1})_{-3}+(\overline{{\bf 10}},{\bf 1},\overline{{\bf 3}})_2 + ({\bf
  5}, {\bf 1},{\bf 1})_{-6} + ({\bf 5},{\bf 1},{\bf 3})_{4} \\
\nonumber &&+ ({\bf 5},{\bf 2},\overline{{\bf 3}})_{-1} +
(\overline{{\bf 5}},{\bf 1},{\bf 1})_6 + (\overline{{\bf 5}},{\bf
  1},\overline{{\bf 3}})_{-4}+ (\overline{{\bf 5}},{\bf 2},{\bf 3})_1
\eea The multiplicities of each representation, as seen in four
dimensions, are given in Table \ref{tableA5}.
\begin{table}[h]
\begin{center}
\begin{tabular}{|c|c|c|}
  \hline
  Representation & Field name & Cohomology \\ \hline
  $({\bf 1},{\bf 2},{\bf 3})_{-5}$ & $C_1$ & $h^1(X,{\cal F}^*\otimes {\cal G})$ \\ \hline
  $({\bf 1},{\bf 2},\overline{{\bf 3}})_{5}$ & $C_2$ & $h^1(X,{\cal F}\otimes {\cal G}^*)$ \\ \hline
  $({\bf 1},{\bf 3},{\bf 1})_{0}$ & $\phi_1$ & $h^1(X,{\cal F} \otimes {\cal F}^*)$ \\ \hline
  $({\bf 1},{\bf 1},{\bf 8})_{0}$ & $\phi_2$ & $h^1(X,{\cal G}\otimes {\cal G}^*)$ \\ \hline
  $({\bf 10},{\bf 2},{\bf 1})_{3}$ & $f_1$ & $h^1(X,{\cal F})$ \\ \hline
  $({\bf 10},{\bf 1},{\bf 3})_{-2}$ & $f_2$ & $h^1(X,{\cal G})$ \\ \hline
  $(\overline{{\bf 10}},{\bf 2},{\bf 1})_{-3}$ & $\tilde{f}_1$ & $h^1(X,{\cal F}^*)$  \\ \hline
  $(\overline{{\bf 10}},{\bf 1},\overline{{\bf 3}})_{2}$ & $\tilde{f}_2$ & $h^1(X,{\cal G}^*)$  \\ \hline
  $({\bf 5},{\bf 1},{\bf 1})_{-6}$ & $h_1$ & $h^1(X,\wedge^2 {\cal F}^*)$ \\ \hline 
  $({\bf 5},{\bf 1},{\bf 3})_{4}$ & $h_2$ & $h^1(X,\wedge^2 {\cal G}^*)$ \\ \hline 
  $({\bf 5},{\bf 2},\overline{{\bf 3}})_{-1}$ & $h_3$ & $h^1(X,{\cal G}^* \otimes {\cal F}^*)$ \\ \hline 
  $(\overline{{\bf 5}},{\bf 1},{\bf 1})_{6}$ & $\tilde{h}_1$ & $h^1(X,\wedge^2 {\cal F})$ \\ \hline 
  $(\overline{{\bf 5}},{\bf 1},\overline{{\bf 3}})_{-4}$ & $\tilde{h}_2$ & $h^1(X,\wedge^2 {\cal G})$ \\ \hline  $(\overline{{\bf 5}},{\bf 2},{\bf 3})_{1}$ & $\tilde{h}_3$ & $h^1(X,{\cal F} \otimes {\cal G})$ \\ \hline 
  \end{tabular}
\mycaption{$SU(5)$ one D-term case 2.}
\label{tableA5}
\end{center}
\end{table}
As always, only one of $C_1$, $C_2$ can have a non-zero vev. This
leads us to two different textures. For $<C_{1}> \neq 0$, we find that

\bea W_{\textnormal{Yukawa}} &=& h_1 f_1^2 + h_2 f_2^2 +h_3 f_1 f_2
+{\bf h_2 f_1^2 + h_2 f_1 f_2 + h_3 f_1^2} \\
\nonumber && + \tilde{h}_1 \tilde{f}_1^2 +  \tilde{h}_2 \tilde{f}_2^2 +\tilde{h}_3 \tilde{f}_1 \tilde{f}_2 +
{\bf \tilde{h}_1 \tilde{f}_1
\tilde{f}_2 + \tilde{h}_1 \tilde{f}_2^2 +
 \tilde{h}_3 \tilde{f}_2^2} \\
\nonumber && +f_1 \tilde{h}_2
\tilde{h}_3 + f_2 \tilde{h}_1 \tilde{h}_2 + f_2 \tilde{h}_3^2+{\bf f_1 \tilde{h}_1^2 + f_1 \tilde{h}_1 \tilde{h}_2 + f_1\tilde{h}_1 \tilde{h}_3 + f_1 \tilde{h}_3^2 + f_2 \tilde{h}_1^2 + f_2
\tilde{h}_1 \tilde{h}_3} \\ 
\nonumber && +
\tilde{f}_1 h_2 h_3 + \tilde{f}_2 h_1 h_2 +{\bf \tilde{f}_1 h_2^2 +  
\tilde{f}_2 h_2^2 + \tilde{f}_2 h_2 h_3 + \tilde{f}_2 h_3^2} \;,\eea
whereas for $<C_{2}> \neq 0$ the following texture
appears

 \bea W_{\textnormal{Yukawa}} &=& h_1 f_1^2 + h_2 f_2^2 +h_3 f_1 f_2
 + {\bf h_1 f_1 f_2 + h_1
f_2^2 +h_3 f_2^2} \\
\nonumber && + \tilde{h}_1 \tilde{f}_1^2 +  \tilde{h}_2 \tilde{f}_2^2 +\tilde{h}_3 \tilde{f}_1 \tilde{f}_2
+{\bf \tilde{h}_1 \tilde{f}_1 \tilde{f}_2 +
\tilde{h}_2 \tilde{f}_1^2 + \tilde{h}_2 \tilde{f}_1 \tilde{f}_2 +
\tilde{h}_3 \tilde{f}_1^2} \\ 
\nonumber && +f_1 \tilde{h}_2 \tilde{h}_3 + f_2 \tilde{h}_1 \tilde{h}_2 +
 f_2 \tilde{h}_3^2+ {\bf f_1 \tilde{h}_2^2 + f_2
\tilde{h}_2^2 + f_2 \tilde{h}_2 \tilde{h}_3} \\
\nonumber && + \tilde{f}_1 h_2 h_3 + \tilde{f}_2 h_1 h_2 +
{\bf \tilde{f}_1 h_1^2 + \tilde{f}_1 h_1 h_2 + \tilde{f}_1 h_3^2 + \tilde{f}_1 h_1 h_3 + \tilde{f}_2 h_1^2
+\tilde{f}_2 h_1 h_3 + \tilde{f}_2 h_3^2} \;.
\eea 
These expressions are grouped by $SU(5)$ products as in the
previous case.

\subsection{Stability Wall Texture: A Three Family Mass Hierarchy}

An important question in string model building is the following:
is there a natural texture in a heterotic vacuum which leads, perturbatively,
to a one heavy and two light families?
To explore this issue, let us consider an $SO(10)$ theory of the type described
in case \ref{su4case2} where $<C_{1}> \neq 0$. Choose the bundle so that the $SO(10)$ non-singlet field
content is  two $f_1$ fields,  one $f_2$ and one $h_1$ only. This is a
three generation model with one pair of Higgs doublets (inside
$h_1$). From \eqref{egtext}, one sees that
the only allowed perturbative Yukawa coupling of this vacuum is
 \bea W_{\textnormal{Yukawa}}= h_1 f_2^2 \ . \eea That is, there is one massive third family and two light
generations as desired. Therefore, one can expect one heavy family to
arise naturally in some smooth compactifications of the heterotic
string.
 
\section{Equivariant structures and quotient manifolds}\label{equivariance}
In this Appendix, we briefly outline the procedure for constructing a vector bundle on a manifold $X/\Gamma$ for some discrete group $\Gamma$. This is intended to provide only a brief introduction to the construction. For a more detailed discussion of building equivariant structures and Wilson line symmetry breaking on non-simply connected Calabi-Yau manifolds see \cite{Anderson:2009mh, Braun:2004xv} and \cite{Gray:2008zs} for useful numeric tools for these calculations.

To begin, we consider a rank $4$ bundle, $V$, on the bicubic hypersurface in $\mathbb{P}^3 \times \mathbb{P}^3$,  
\beq\label{33}
X= \left[\begin{array}[c]{c}\mathbb{P}^2\\\mathbb{P}^2\end{array}
\left|\begin{array}[c]{ccc}3 \\3
\end{array}
\right.  \right]^{2,83}\;
\eeq
 If we denote the coordinates on $\mathbb{P}^2\times \mathbb{P}^{2}$ by $\{x_i,y_i\}$, where $i=0,1,2$, then a freely acting $\mathbb{Z}_3 \times \mathbb{Z}_3$ symmetry is generated by \cite{Candelas:2007ac},
\bea\label{z3z3}
{\mathbb{Z}_3}^{(1)}~~~~:~~~~~~x_k \to x_{k+1},~~~y_k \to y_{k+1}~~~~~~~~\\ \nonumber
{\mathbb{Z}_3}^{(2)}~~~~:~~~~x_k \to e^{\frac{2\pi ik}{3}} x_{k},~~~y_k \to e^{-\frac{2\pi ik}{3}} y_{k}~.
\eea

As shown in \cite{Candelas:2007ac}, the most general bi-degree $\{3,3\}$ polynomial invariant under the above symmetry is given by
\beq\label{bicubic_poly}
p_{\{3,3\}}=A_{1}^{k,\pm} \sum_{j} x_{j}^{2}x_{j\pm 1}y^{2}_{j+k}y_{j+k\pm1}+A_{2}^{k}\sum_{j}x^{3}_{j}y^{3}_{j+k}+A_{3}x_1x_2x_3\sum_{j}y^{3}_j+A_{4}y_1y_2y_3\sum_{j}x^{3}_{j}+A_5x_1x_2x_3y_1y_2y_3
\eeq
where $j,k=0,1,2$ and there are a total of $12$ free coefficients, $A$, above. We shall take these coefficients to be generic (i.e. random integers) in the following. With the choice of the invariant polynomial \eref{bicubic_poly}, we can define the smooth quotient manifold $\hat{X}=X/G$.

The quotient manifold is related to $X$ via the natural projection map, $q: X \rightarrow \hat{X}$. Using this relationship, we note that any vector bundle $\hat{V}$ on $\hat{X}$ can be related to a vector bundle $V$ on $X$ via the pullback map, $q^*$. That is, for any bundle $\hat{V}$ over $\hat{X}$,
\beq
q^*(\hat{V}) \simeq V
\eeq
is an isomorphism for some bundle $V$ on $X$. This pulled-back bundle, $V$ is characterized by the geometric property of {\it equivariance}. A vector bundle on $X$ will descend to a bundle $\hat{V}$ on $\hat{X}$ if for each element $g \in G$, $g: X \to X$, there exists a bundle isomorphism $\phi_g$, satisfying two properties. First, $\phi_g$ must cover the action of $g$ on $X$ such that the following diagram commutes
\begin{equation}
  \begin{array}{lllll}
  &V&\stackrel{\phi_g}{\longrightarrow}&V&\\
  \pi &\downarrow&&\downarrow&\pi\\
  &X&\stackrel{g}{\longrightarrow}&X&
 \end{array}
 \label{invalt}
\end{equation} 
and in addition, $\phi_g$ must satisfy a so-called {\em co-cycle
condition}, namely that for all $g,h \in G$, 
\beq\label{cocycle} \phi_g
\circ \phi_h =\phi_{gh} \;. \eeq
The set of such isomorphisms $\phi_g$ are referred to collectively as an {\it equivariant structure}. The morphisms $\phi_g$ form a representation of the group that act on the bundle, the so-called lifting of \eref{z3z3} to $V$. In addition, $\phi$ induces a representation of the group that acts on the cohomology $H^i(X, V)$ and it is precisely the {\it invariant} elements of this cohomology (under the action inferred from $\phi$) that descend to the quotient manifold $X/\Gamma$. We will return to this later. For now, we consider the description of the bundle $V$ on $X$.

\subsection{The ``upstairs" theory}
For our current purposes, we shall choose the following equivariant bundle $V$ on $X$. The bundle is defined by the following short exact sequence:
\beq\label{pheno_eg}
0 \to \cF_1 \to V \to \cF_2 \to 0
\eeq
where 
\bea\label{fdefs}
0 \to \cF_1 \to \cO(0,2)^{\oplus 3} \to \cO(2,2) \to 0\\
0 \to \cF_2 \to \cO(1,-1)^{\oplus 3} \to \cO(1,1) \to 0 \;.
\eea
The bundle $V$ is defined as an extension of $\cF_1$ by $\cF_2$, where $\cF_i$ are rank 2 bundles defined by monad sequences. Since $c_1(\cF_1)=(-2,4)$ and $c_1(\cF_2)=(2,-4)$, we see that $c_1(V)=0$ and $V$ is a an $SU(4)$ bundle. 

To analyze the properties of the four-dimensional effective theory associated to $\hat{V}$ (including the stability-wall induced textures in its Yukawa couplings) we must first consider the ``upstairs" bundle $V$ and use its properties to determine those of $\hat{V}$, ``downstairs". To begin then, the ``upstairs" spectrum of $V$ is given by
\beq\label{spec_up}
\ba
{l}
n_{16}=h^1(X,V)=h^1(X,\cF_1)+h^1(X,\cF_2)=9+18=27 \\
n_{\bar{16}}=h^1(X,V^*)=0 \\
n_{10}=h^1(X,\wedge^2 V)=h^1(X,\wedge^2 \cF_1)=18
\ea
\eeq

A simple analysis along the lines of \cite{Anderson:2009nt,stability_paper}, shows us that $\cF_1$ and $\cF_2$ are both stable independently, but that $\cF_1 \in V$ itself destabilizes $V$ in a part of its K\"ahler cone. Since $c_1(\cF_1)=(-2,4)$, we see that there is a stability wall when $t^2/t^1=1+\sqrt{3}$. We find that the regions of stability for $V$ on $X$ are as shown in Figure \ref{so10up}. 

At this stability wall, the poly-stable decomposition of $V$ is as the direct sum of the two rank $2$, bundles, $V \rightarrow \cF_1 \oplus \cF_2$. As we have argued in the previous sections, this supersymmetric decomposition of $V$ changes the structure group of the bundle from $SU(4)$ to $S[U(2)\times U(2)]$ and hence, an extra $U(1)$ appears in the low energy gauge symmetry. At this locus in moduli space, the visible matter fields, the ${\bf 16}$s and ${\bf 10}$s in \eref{spec_up}, carry a charge under the enhanced $U(1)$ as shown in Table \ref{tableupstairs} (the general case of such a decomposition is given in Table \ref{tableA3} in Appendix A). 

\begin{table}[h]
\begin{center}
\begin{tabular}{|c|c|c||c|}
  \hline
  Representation & Field name & Cohomology& Multiplicity \\ \hline
  $({\bf 1},{\bf 2},{\bf 2})_{2}$ & $C_1$ & $H^1(X,{\cal F}_1\otimes {\cal F}_2^*)$& 90 \\ \hline
   $({\bf 1},{\bf 3},{\bf 1})_{0}$ & $\phi_1$ & $H^1(X,{\cal F}_1 \otimes {\cal F}_1^*)$& 9 \\ \hline
  $({\bf 1},{\bf 1},{\bf 3})_{0}$ & $\phi_2$ & $H^1(X,{\cal F}_2\otimes {\cal F}_2^*)$& 9 \\ \hline
  $({\bf 16},{\bf 2},{\bf 1})_{1}$ & $f_1$ & $H^1(X,{\cal F}_1)$& 18 \\ \hline
  $({\bf 16},{\bf 1},{\bf 2})_{-1}$ & $f_2$ & $H^1(X,{\cal F}_2)$& 9 \\ \hline
 $({\bf 10},{\bf 1},{\bf 1})_{2}$ & $h_1$ & $H^1(X,\wedge^2 {\cal F}_1)$&18 \\ \hline 
  \end{tabular}
\mycaption{The ``upstairs" field content of the decomposed bundle $V\rightarrow \cF_1 \oplus \cF_2$.}
\label{tableupstairs}
\end{center}
\end{table}

\begin{figure}[!ht]
  \centerline{\epsfxsize=3.5in\epsfbox{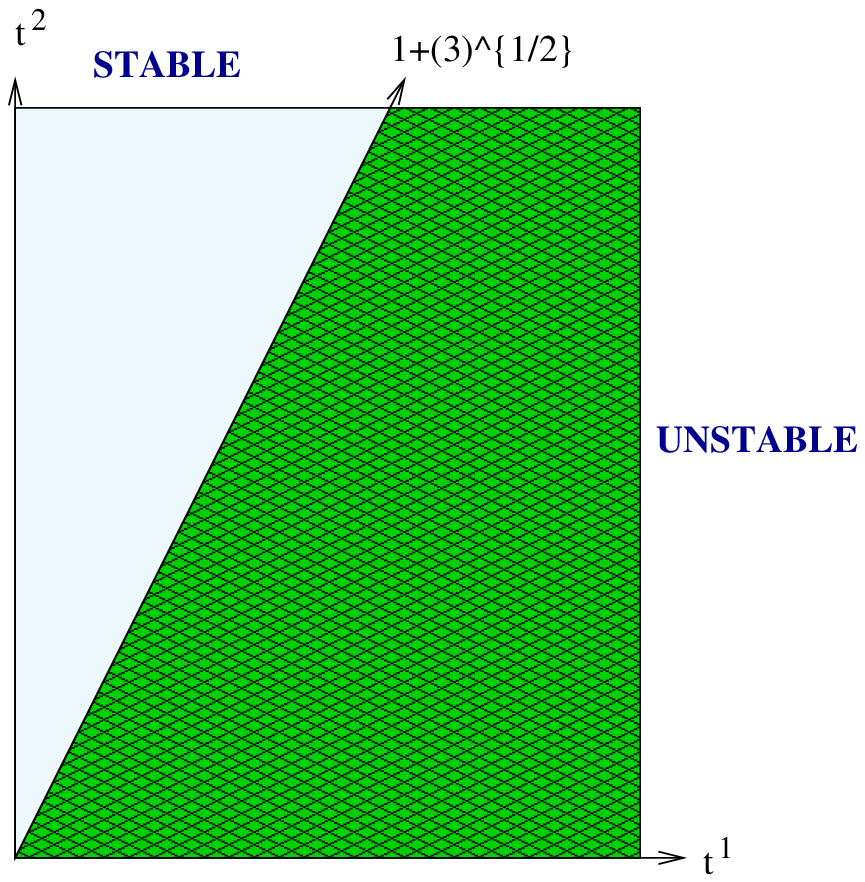}} \mycaption{
    The K\"ahler cone ($t^1,t^2>0$) and the regions of stability/instability for the ``upstairs" bundle \eqref{pheno_eg} on the simply connected Calabi-Yau \eqref{33}. At the stability wall ($t^2/t^1=1+\sqrt{3}$), $V$ decomposes as $V \rightarrow \cF_1 \oplus \cF_2$ where $\cF_1$ and $\cF_2$ are defined in \eref{fdefs}.}
\label{so10up}
\end{figure}

\subsection{The ``downstairs" theory}
We turn now to the final, three generation theory on the quotient manifold $\hat{X}=X/\mathbb{Z}_3\times \mathbb{Z}_3$. By quotienting the Calabi-Yau threefold in \eref{33} by the discrete symmetry in \eref{z3z3}, we form the manifold
\beq\label{33quad}
X= \left[\begin{array}[c]{c}\mathbb{P}^2\\\mathbb{P}^2\end{array}
\left|\begin{array}[c]{ccc}3 \\3
\end{array}
\right.  \right]^{2,11}_{/\mathbb{Z}_3\times \mathbb{Z}_3}
\eeq

Since each term in the short exact sequence \eref{pheno_eg} is equivariant, the entire sequence descends to a sequence of bundles over $\hat{X}$. We have
\beq
0 \to \hat{\cF_1} \to \hat{V} \to \hat{\cF_2} \to 0
\eeq
Using the fact that the cohomology of $\hat{V}$ is simply the invariant part of the cohomology $V$ under the induced action of $\phi_g$ in \eref{invalt} and \eref{cocycle}, we find that the number of ${\bf 16}$s and ${\bf 10}$s of $SO(10)$ are
\beq
\ba
{l}
f_1: h^1(\hat{X},\hat{F}_{2})=2 \\
f_2: h^1(\hat{X},\hat{F}_{1})=1\\
h_1: h^1(\hat{X},\wedge^2 \hat{F}_2)=2
\ea
\eeq
and that the charged bundle moduli $C_1$ become
\beq
C_1: h^1(\hat{X},\hat{\cF_1}\times\hat{\cF_2}^*)=9
\eeq

From the above, it is clear that we have produced a three generation $SO(10)$ GUT theory. However, we can go still further by introducing Wilson lines which will break $SO(10)$ to $SU(3)\times SU(2) \times U(1)_{Y} \times U(1)_{B-L}$. We shall not go into this breaking here, but refer the reader to \cite{Braun:2004xv,Braun:2005zv,Anderson:2009mh} for details of breaking $SO(10)$ with $\mathbb{Z}_3 \times \mathbb{Z}_3$ Wilson lines.


\end{document}